\documentclass[a4paper,11pt]{article}
\pdfoutput=1 

\usepackage{jheppub_mod} 

\usepackage[T1]{fontenc} 

\usepackage{cancel}

\usepackage{enumitem}
\usepackage{mathrsfs}  
\usepackage{mathtools}

\newcommand{\BB}{{\scriptscriptstyle\text{B}}}
\newcommand{\FF}{{\scriptscriptstyle\text{F}}}

\usepackage{tikz}
\usetikzlibrary{math}
\usetikzlibrary{arrows,shapes,positioning}
\usetikzlibrary{decorations.markings}
\tikzstyle arrowstyle=[scale=1]
\tikzstyle directed=[postaction={decorate,decoration={markings,
    mark=at position .65 with {\arrow[arrowstyle]{stealth}}}}]
\tikzstyle reverse directed=[postaction={decorate,decoration={markings,
    mark=at position .65 with {\arrowreversed[arrowstyle]{stealth};}}}]

\newlength{\mywidth}
\usepackage[colorlinks=true
,urlcolor=blue
,anchorcolor=blue
,citecolor=blue
,filecolor=blue
,linkcolor=blue
,menucolor=blue
,linktocpage=true
,pdfproducer=medialab
,pdfa=true
]{hyperref}

\newcommand{\bei}{\begin{itemize}}
\newcommand{\eei}{\end{itemize}}
\newcommand{\bee}{\begin{enumerate}}
\newcommand{\eee}{\end{enumerate}}
\newcommand{\beeL}{\begin{enumerate}[label=(\Alph*)]}
\newcommand{\beel}{\begin{enumerate}[label=(\alph*)]}
\newcommand{\beeR}{\begin{enumerate}[label=(\Roman*)]}
\newcommand{\beer}{\begin{enumerate}[label=(\roman*)]}
\newcommand{\beeLd}{\begin{enumerate}[label=\Alph*.]}
\newcommand{\beeld}{\begin{enumerate}[label=\alph*.]}
\newcommand{\beeRd}{\begin{enumerate}[label=\Roman*.]}
\newcommand{\beerd}{\begin{enumerate}[label=\roman*.]}

\newcommand{\bal}{\begin{equation}\begin{aligned}}
\newcommand{\eal}{\end{aligned}\end{equation}}
\newcommand{\ov}{\over}
\newcommand{\g}{\gamma}

\newcommand{\lint}{\int\limits}

\def\sgn{{\text{sgn}}}

\definecolor{grey}{rgb}{0.4,0.4,0.5}
\definecolor{darkgreen}{rgb}{0,0.5,0}
\definecolor{darkred}{rgb}{0.6,0.0,0}
\definecolor{lightbrown}{rgb}{1,0.9,0.8}
\definecolor{brown}{rgb}{0.6,0.3,0.3}
\definecolor{darkblue}{rgb}{0,0,0.5}
\definecolor{darkmagenta}{rgb}{0.5,0,0.5}

\def\tx{{\tilde x}}

\def\tg{{\tilde\gamma}}

\def\tPhi{{\tilde\Phi}}

\def\a {\alpha}

\def\bes{{\text{\tiny BES}}}
\def\afs{{\text{\tiny AFS}}}
\def\hl{{\text{\tiny HL}}}
\def\tp{{\widetilde p}}
\def\tE{\widetilde\E}

\newcommand{\E}{\mathcal E}

\newcommand{\B}{{\scriptscriptstyle\text{B}}}
\newcommand{\F}{{\scriptscriptstyle\text{F}}}
\renewcommand{\L}{{\scriptscriptstyle\text{L}}}
\newcommand{\R}{{\scriptscriptstyle\text{R}}}

\def\ka {{\kappa}}

\def\pa {\partial}

\def\cA{{\cal A}}
\def\cB{{\cal B}}

\def\cO{{\cal O}}

\def\cR{{\cal R}}

\def\cX{{\cal X}}

\def\cZ{{\cal Z}}

\newcommand{\de}{\text{d}}

\newcommand{\AdSSSS}{\text{AdS}_3\times\text{S}^3\times\text{S}^3\times\text{S}^1}
\newcommand{\AdSST}{\text{AdS}_3\times\text{S}^3\times\text{T}^4}

\title{\boldmath On the AdS$_3 \, \times$ S$^3 \, \times$ S$^3 \, \times$ S$^1$ dressing factors}

\author[a]{Sergey Frolov,}
\author[b,c,1]{Alessandro Sfondrini\note{On leave from the University of Padova, Italy.}}

\affiliation[a]{Hamilton Mathematics Institute and School of Mathematics Trinity College, Dublin 2, Ireland.}
\affiliation[b]{School of Mathematics, University of Birmingham, Edgbaston B15 2TT, UK.}
\affiliation[c]{Istituto Nazionale di Fisica Nucleare, Sezione di Padova, via Marzolo 8, 35131 Padova,
Italy.}

\emailAdd{frolovs@maths.tcd.ie}
\emailAdd{a.sfondrini@bham.ac.uk}

\abstract{We propose dressing factors for massive excitations of the worldsheet S matrix of $\AdSSSS$ supported by mixed Ramond--Ramond and Neveu-Schwarz--Neveu-Schwarz flux, in the ``string'' and ``mirror'' kinematics.
Our proposal is compatible with crossing and unitarity, and it reproduces the available perturbative results for any ratio of the two three-spheres' radii.
}

\begin{document} 
\maketitle
\flushbottom

\section{Introduction}
The $\AdSSSS$ geometry provides a rich playground for string theory and holography, see e.g.~\cite{Maldacena:1997re,Elitzur:1998mm,Boonstra:1998yu,Babichenko:2009dk,Cagnazzo:2012se,Tong:2014yna,Borsato:2015mma,Eberhardt:2017pty,Witten:2024yod}. The relative size of the two spheres is not fixed by the supergravity equations and the background enjoys 16 Killing spinors (the maximal amount for an AdS$_3$ background) regardless of the ratio of the spheres' radii.
In fact, the supergravity equations only require the curvature radii of AdS and of the two three-spheres to satisfy
\begin{equation}
\frac{1}{R_{\text{AdS}_3}^2} = \frac{1}{R_{\text{S}^3_{(1)}}^2} + \frac{1}{R_{\text{S}^3_{(2)}}^2} \,.
\end{equation}
This gives a one-parameter family of backgrounds, which we may label by $\alpha$. 
Setting for convenience $R_{\text{AdS}_3}^2=1$, one defines
\begin{equation}
\frac{1}{R_{\text{S}^3_{(1)}}^2}= \alpha \,, \qquad \frac{1}{R_{\text{S}^3_{(2)}}^2}=1-\alpha\,, \qquad 0 < \alpha < 1\,.
\end{equation}
The 16 Killing spinors organise themselves in two copies of the exceptional Lie superalgebra $\mathfrak{d}(2,1;\alpha)$. 
In the limiting cases $\alpha\to0$ or $\alpha\to1$, either sphere decompactifies and the superalgebra contracts to $\mathfrak{psu}(1,1|2)$ up to Abelian factors.%
\footnote{For $\alpha=1/2$, the exceptional Lie superalgebra $\mathfrak{d}(2,1;\alpha)$ is isomorphic to the orthosymplectic superalgebra $\mathfrak{d}(2,1)$,  also known as $\mathfrak{osp}(4|2)$, see~\cite{Frappat:1996pb}. This is the case where the spheres have the same radius.}

On top of the parameter $\alpha$, any such $\AdSSSS$ geometry can be supported by NSNS three-form flux, RR flux, or a mixture of the two, without breaking any supersymmetry. Like it happens in the simpler $\AdSST$ case, this allows for another parameter which encodes the relative strength of RR and NSNS fluxes.  Different choices of fluxes may be related by S~duality, but in perturbative string theory it is natural to consider them as distinct, since they have rather different features (and spectrum).
Writing the bosonic piece of the string action in the mixed-flux case as
\begin{equation}
S_{\text{bos}}=-\frac{T}{2}\int\limits_{-\infty}^{+\infty}\de\tau\int\limits_0^{R}\de\sigma\left[\gamma^{ij}G_{\mu\nu}(X)+\varepsilon^{ij}B_{\mu\nu}(X)\right]\partial_iX^\mu\partial_j X^\nu\,,
\end{equation}
where $\gamma_{ij}$ is the unit-determinant worldsheet metric, $R$ is the volume of the worldsheet theory (the length of the string), and $T$ is string tension, one finds that (in the notation of~\cite{Borsato:2015mma})  the NSNS flux is given by
\begin{equation}
    H=\de B=2q\,\left[\text{vol}(\text{AdS}_3)+\frac{1}{\alpha}\text{vol}(\text{S}^3_{(1)})+\frac{1}{1-\alpha}\text{vol}(\text{S}^3_{(2)})\right]\,,\qquad 0\leq q\leq1\,,
\end{equation}
where `vol' denotes the volume form and $q$ interpolates between the pure-RR case ($q=0$) and the pure-NSNS one ($q=1$). The quantisation condition for the spheres dictates that
\begin{equation}
    \frac{k_1}{2\pi}= \frac{Tq}{\alpha}\,,\qquad
    \frac{k_2}{2\pi}=\frac{Tq}{1-\alpha}\,,\qquad k_j\in\mathbb{N}\,,\quad j=1,2\,,
\end{equation}
which in analogy with $\AdSST$ can be rewritten as
\begin{equation}
    T=\sqrt{h^2+\frac{k^2}{4\pi^2}}\,,\qquad k=\frac{k_1\,k_2}{k_1+k_2}\,,\qquad\alpha=\frac{k_2}{k_1+k_2}\,,
\end{equation}
where $h\geq0$ is related to the amount of RR flux.\footnote{More precisely, at large tension $T\gg1$, the expression $h=\sqrt{1-q^2}\,T+\mathcal{O}(T^0)$ appears in form of the AdS volume form in the RR-three form.}
Note that while $k_1,k_2$ are integers, $k$ is not necessarily so, and that the limit in which the sphere $S^{3}_{(j)}$ blows up is now $k_j\to\infty$, with $j=1,2$.

In the special case $q=1$ (whereby $h=0$), the worldsheet action in conformal gauge becomes that of a Wess-Zumino-Witten model based on the $sl(2)_{k}\oplus su(2)_{k_1}\oplus su(2)_{k_2}$ Ka\v{c}-Moody algebra. In this case, the superstring can be efficiently studied in the RNS formalism~\cite{Maldacena:2000hw}. As usual, this becomes much harder in the presence of RR fluxes.
Remarkably, for any value of $k_1, k_2$ and $h$, the Green-Schwarz action is classically integrable~\cite{Babichenko:2009dk,Cagnazzo:2012se}. This paved the way to an ambitious program to solve the mixed-flux superstring model by worldsheet integrability~\cite{OhlssonSax:2011ms,Borsato:2012ud,Borsato:2012ss,Borsato:2015mma}, see~\cite{Sfondrini:2014via,Demulder:2023bux} for reviews of this progress in the context of $\text{AdS}_3$ superstrings.

Let us briefly recall the main steps of the worldsheet-integrability approach~\cite{Arutyunov:2009ga,Beisert:2010jr}. Starting from the Green--Schwarz action in the ``uniform'' lightcone gauge~\cite{Arutyunov:2005hd} and large volume $R\to\infty$, one obtains a worldsheet theory whose free Hamiltonian coincides with the pp-wave one, see~\cite{Rughoonauth:2012qd,Sundin:2013ypa}. Beyond this order, one can compute perturbatively (in $T\gg1$) the scattering matrix of worldsheet excitations, though for AdS$_3$ backgrounds the computation is riddled with subtleties (both infrared and ultraviolet divergences). In practice, one may trust the tree-level results and the piece of one-loop results which is cut-constructible with unitarity techniques. As it turns out, it is actually possible to \textit{bootstrap} the worldsheet S~matrix instead of computing it in perturbation theory. For $\AdSSSS$, this was done in~\cite{Borsato:2015mma} in the same spirit as what had been done for $\text{AdS}_5\times \text{S}^5$ in~\cite{Arutyunov:2006ak,Arutyunov:2006yd}.%
\footnote{The $\AdSSSS$ S~matrix was first studied for RR backgrounds in~\cite{Borsato:2012ud}, and an important analysis of the worldsheet symmetries was carried out in~\cite{Hoare:2013lja}, laying the groundwork for the study of the full mixed-flux S~matrix in~\cite{Borsato:2015mma}.}
The symmetries surviving the lightcone gauge fixing acquire a central extension which acts non-trivially on states not satisfying the level-matching constraint. Unlike the case of $\AdSST$ and of $\text{AdS}_5\times \text{S}^5$, for $\AdSSSS$ only \textit{one quarter} of the supersymmetries survive lightcone gauge fixing (rather than one half).
Worldsheet excitations transform in short representations of the lightcone symmetry algebra. These are two-dimensional, and consist of a boson and a fermion; we will briefly review them below. Each S-matrix block scattering these representations is fixed by the lightcone symmetries \textit{up to an overall undetermined scalar factor} --- the dressing factor.%
\footnote{Because the irreducible representations have smaller dimension for $\AdSSSS$ than for $\AdSST$, there are more dressing factors to be determined in the former case.}
This is where the investigation of strings on mixed-flux $\AdSSSS$ backgrounds currently stands.%
\footnote{The case of pure-NSNS backgrounds is especially simple, and in that case a proposal for the S-matrix, dressing factors and mirror TBA was put forward in~\cite{Dei:2018jyj} and checked against the WZW construction.}
In general, dressing factors can be fixed by demanding invariance under crossing, unitarity, and ``good analytic properties''. 
Once  the S matrix is completely known, one considers a ``mirror model''~\cite{Arutyunov:2007tc}, related to the original worldsheet model by double Wick rotation
\begin{equation}
    \tau =i\,\tilde{\sigma}\,,\qquad
    \sigma=i\,\tilde{\tau}\,.
\end{equation}
The thermodynamic Bethe ansatz (TBA) for the mirror model then allows one to find the finite-volume  spectrum of the original string model by means of the contour deformation trick~\cite{Zamolodchikov:1989cf,Dorey:1996re,Arutyunov:2009ax}. In practice, original ``string'' model and the new ``mirror'' model are related by an analytic continuation, as first described in~\cite{Arutyunov:2007tc} for $\text{AdS}_5\times \text{S}^5$ superstrings and in~\cite{Frolov:2025uwz,Frolov:2025tda} for mixed-flux $\AdSST$ superstrings.  In the case at hand, the kinematics of the string and mirror models is closely related to the one of mixed-flux $\AdSST$.

We should mention that another approach to computing the string spectrum from integrability has recently been taken. From the mirror TBA, it is possible to derive a set of ``quantum spectral curve'' (QSC) equations (similar to QQ relations) which is in principle equivalent but in practice often computationally more powerful. Because the QSC must satisfy rather strict self-consistency conditions, it may be possible to make an educated guess for its form just based on the symmetries and particle content of model under consideration. For pure-RR $\AdSSSS$ backgrounds this was very recently done in~\cite{Cavaglia:2025icd,Chernikov:2025jko}, especially in the case where $\alpha=1/2$.
Because the QSC is in principle equivalent to the S-matrix and mirror-TBA construction outlined above, it should be possible to recover from the QSC the dressing factors, though this is not always transparent.

The aim of this paper is to complete the S-matrix bootstrap for string of mixed-flux $\AdSSSS$ (for arbitrary $\alpha$) by proposing dressing factors which solve the crossing and unitarity constraints, match the known perturbative data, and have consistent analytic properties. We will focus on the scattering of massive excitations; we expect the dressing factors involving massless modes to be obtainable by a suitable limiting procedure, like in~\cite{Frolov:2025tda}.
We will also comment on how our proposal for the dressing factors compares with the QSC proposals~\cite{Cavaglia:2025icd,Chernikov:2025jko}, which were put forward for the pure-RR case ($k_1=k_2=0$) and mostly concerned with $\alpha=1/2$.

This article is structured as follows. In section~\ref{sec:ws} we recall some of the key features of the worldsheet model, including its crossing equations, and we comment on the  expected pole structure of the S~matrix. In section~\ref{sec:proposal} we put forward our proposal for the massive dressing factors and we discuss the resulting Bethe strings in the mirror kinematics. In section~\ref{sec:conclusions} we discuss some open questions as well as the relation of our proposal to the recently-proposed QSC.  In the appendices, we summarise some useful formulae (appendix~\ref{app:def}), prove unitarity and crossing (appendix~\ref{app:crossing}) and check the near-BMN expansion (appendix~\ref{app:BMN}). Finally, in appendix~\ref{app:solutions} we discuss two more solutions which can be found in the case $\alpha=1/2$; while they appear less physical than the one presented in the main text, it is nonetheless interesting to note their existence.
 
\section{The worldsheet model}
\label{sec:ws}

The lightcone-gauge fixed model was studied in detail in~\cite{Borsato:2015mma}. As usual, after lightcone- and $\kappa$-gauge fixing, we are left with eight real bosons and fermions. Two of the bosons are related to the transverse modes on $\text{AdS}_3$, and we denoted them by $Z,\bar{Z}$; for the first and second sphere we have $Y,\bar{Y}$ and $X,\bar{X}$, respectively. This notation highlights the fact that each pair of bosons has opposite charge under a $\mathfrak{u}(1)$ symmetry --- the spin in AdS$_3$, or the spin on either three-sphere, respectively. The remaining two bosons are called $T, \bar{T}$. They are both massless and are related to the S$^1$ direction and to a linear combination of  coordinates  on the equator of the three spheres which is orthogonal to the one used in the gauge fixing, see e.g.~\cite{Dei:2018yth} for a detailed discussion.

\subsection{Lightcone symmetries and  representations}

Because the lightcone gauge fixing preserves only one quarter of the 16 original Killing spinors, the lightcone supersymmetry algebra of $\AdSSSS$ is given by
\begin{equation}
\label{eq:lcalgebra}
\begin{aligned}
    &\{ \mathbf{Q},  \mathbf{S} \}= \frac{1}{2} (\mathbf{H}+\mathbf{M})\,, \qquad&\{ \mathbf{Q},  \mathbf{\tilde{Q}} \}= \mathbf{C} \, ,&\\
    &\{ \mathbf{\tilde{Q}},  \mathbf{\tilde{S}} \}= \frac{1}{2} (\mathbf{H}-\mathbf{M}), \qquad&\{ \mathbf{S},  \mathbf{\tilde{S}} \}= \mathbf{\bar{C}} \, .&
\end{aligned}
\end{equation}
Here $\mathbf{H}$ is the lightcone Hamiltonian, $\mathbf{M}$ is a linear combination of the spins on AdS$_3$ and on the two three-spheres, and $\mathbf{C},\mathbf{\bar{C}}$ are the off-shell central extensions, see~\cite{Arutyunov:2006ak,Borsato:2015mma}.
The operators $\mathbf{H}$, $\mathbf{M}$, $\mathbf{C}$ and $\mathbf{\bar{C}}$ are diagonal operators on single-particle states having as eigenvalues:
\begin{equation}
\label{eq:representation-charges}
\begin{aligned}
        &H(p)=\sqrt{\Bigl(\mu+\frac{k}{2 \pi}p \Bigr)^2 +4 h^2 \sin^2 \frac{p}{2}} \, ,\qquad&&M(p)=\mu+\frac{k}{2 \pi}p \, ,\\
        &C(p)= \frac{i h}{2} \bigl( e^{ip}-1 \bigr) \, ,\qquad
        &&\bar{C}(p)= -\frac{i h}{2} \bigl( e^{-ip}-1 \bigr) \, .
        \end{aligned}
\end{equation}
Here $p$ is the momentum of the particle, while $\mu$ labels different representations, so that $m=|\mu|$ can be interpreted as a mass of sorts. Its values on the bosons are summarised in Table~\ref{tab:masses}.
Here we are mainly interested in the values $m=|\mu|=\alpha$ and $m=|\mu|= 1- \alpha$. These are the lightest massive particles of the theory. As we will discuss, heavier excitations may arise of bound states of these, and the allowed bound-state masses may be different in the string and mirror kinematics. Suggestively, the masses of the modes on either sphere add up to give precisely mass of the  $AdS_3$ modes. For this reason, one may expect that $Z,\bar{Z}$ could be bound-states or composite modes of the lighter excitations. It has been conjectured that the latter is the case~\cite{Sundin:2012gc} in analogy with what happens for strings on $\text{AdS}_4\times \mathbb{C}\text{P}^3$~\cite{Zarembo:2009au}.

\begin{table}[t]
\centering
\begin{tabular}{c|cc|cc|cc|cc}
Boson & $Z$ & $\bar{Z}$  &  $Y$ & $\bar{Y}$  &$X$ & $\bar{X}$  &$T$ & $\bar{T}$  \\ \hline
$\mu=$ & $+1$ & $-1$ & $+\alpha$ & $-\alpha$  & $+1-\alpha$ & $-1+\alpha$& 0 &0
\end{tabular}
\label{tab:masses}
\caption{Values of~$\mu$ in~\eqref{eq:representation-charges} on the bosons appearing in the near-pp-wave expansion. In what follows, we will often use $m=|\mu|$ to identify the mass of the particles.}
\end{table}

These bosonic excitations combine with a fermionic one to make up a two-dimensional short irreducible representation of the algebra~\eqref{eq:lcalgebra}.
For these representations, we introduce the notation $\rho_\star^\BB(m,p)$, in the case the highest weight state (HWS) is  \textit{bosonic};%
\footnote{In our convention, $\mathbf{Q}$ and $\tilde{\mathbf{S}}$ are lowering operators, while $\tilde{\mathbf{Q}}$ and $\mathbf{S}$ are raising operators.}
$m=|\mu|$ is the mass and $p$ is the momentum. Similarly we introduce a representation $\rho_\star^\FF(m,p)$, having a \textit{fermionic} highest weight state (HWS), mass $m=|\mu|$ and momentum $p$. The placeholder $\star$ can either be L (left) when $\mu>0$,  or R (right) when $\mu<0$.
As explained in~\cite{Borsato:2015mma} the left- and right-representations can be related when $\mu=0$. 
The L and R labels are reflected in the parameterisation used for the representation in terms of Zhukovsky variables (see appendix~\ref{app:def}) and plays an important role for the crossing equations of this model. In particular, crossing maps L and R representations (and Zhukovsky variables) to each other as discussed in~\cite{Lloyd:2014bsa}. 
We label particle short-representations by $(\cdot, \cdot)$, where the first entry  always corresponds to the HWS and the second to the LWS; we denote the fermions by Greek letters.

\paragraph{Particles of mass $\alpha$.}

Left and right particles of mass $\alpha$ are in the representations
\bal
(Y, \psi) \in \rho_{\L}^{\B}\,, \qquad (\bar{\psi}, \bar{Y}) \in \rho_{\R}^{\F} \,.
\eal

\paragraph{Particles of mass $1-\alpha$.}

Left and right particles of mass $1-\alpha$ are in the representations
\bal
(X, \chi) \in \rho_{\L}^{\B}\,, \qquad (\bar{\chi}, \bar{X}) \in \rho_{\R}^{\F} \,.
\eal

\paragraph{Particles of mass $0$.}

They are in the representations
\bal
(T, \zeta) \in \rho_{\L}^{\B}\,, \qquad (\bar{\zeta}, \bar{T}) \in \rho_{\R}^{\F} \,.
\eal

\paragraph{Particles of mass $1$.}

We expect particles  of mass $1$ to be composite, therefore we will not discuss them below. They can be fit in the representations
\bal
(\vartheta, Z) \in \rho_{\L}^{\F}\,, \qquad (\bar{Z}, \bar{\vartheta}) \in \rho_{\R}^{\B} \,.
\eal

.

\subsection{Crossing equations}
We write the crossing equations in terms of the whole S-matrix elements (without stripping out any prefactor), in the string and mirror kinematics (denoting the latter with tildes). Rather than expressing them in terms of the momentum, we use the rapidity variable~$u$ and the Zhukovsky variables defined in appendix~\ref{app:def}.
In particular, we are interested in the following crossing equations:
\begin{equation}
\label{eq:crossinghw}
\begin{aligned}
S^{m_1 m_2}_{\mathcal{X}\mathcal{X}}(u_1, u_2) \,S_{\bar{\Upsilon}\mathcal{X} }^{m_1 m_2}(\bar{u}_1, u_2)=& +f(x_{\L1}^{\pm m_1},x_{\L2}^{\pm m_2}),
\\
S_{\bar{\Upsilon}\mathcal{X} }^{m_1 m_2}({u}_1, u_2) \,S^{m_1 m_2}_{\mathcal{X}\mathcal{X}}(\bar{u}_1, u_2)  =&+g(x_{\R1}^{\pm m_1},x_{\L2}^{\pm m_2}),
\\
S^{m_1 m_2}_{\Upsilon \Upsilon}(u_1, u_2) \,S_{\bar{\mathcal{X}}\Upsilon }^{m_1 m_2}(\bar{u}_1, u_2)=&-\tilde{f}(\tx_{\L1}^{\pm m_1},\tx_{\L2}^{\pm m_2}),
\\
S_{\bar{\mathcal{X}}\Upsilon }^{m_1 m_2}(u_1, u_2)\, S^{m_1 m_2}_{\Upsilon \Upsilon}(\bar{u}_1, u_2) =&-\tilde{g}(\tx_{\R1}^{\pm m_1},\tx_{\L2}^{\pm m_2}),
\end{aligned}
\end{equation}
where $\bar{u}_1$ has the same numerical value as $u_1$ but it is reached by crossing the string or mirror theory main cuts of both $\tx^+_1$ and $\tx^-_1$. We also use the notation
\begin{equation}
    \mathcal{X}=X,Y,T,\qquad
    \bar{\mathcal{X}}=\bar{X},\bar{Y},\bar{T},\qquad
    \Upsilon=\psi,\chi,\zeta,\qquad
    \bar{\Upsilon}=\bar{\psi},\bar{\chi},\bar{\zeta},
\end{equation}
so that for instance 
\begin{equation}
    S^{\alpha,\alpha}_{\Upsilon \Upsilon}(u_1, u_2)\equiv S^{\alpha,\alpha}_{\psi\psi}(u_1, u_2)\,,\qquad
    S^{\alpha,1-\alpha}_{\Upsilon \Upsilon}(u_1, u_2)\equiv S^{\alpha,1-\alpha}_{\psi\chi}(u_1, u_2)\,,\qquad\dots\,.
\end{equation}
Four more equations, on top of~\eqref{eq:crossinghw}, may be obtained by left-right symmetry, i.e.\ considering the ``barred'' excitations and swapping~L$\leftrightarrow$R.
The explicit minus sign in the last two equations of~\eqref{eq:crossinghw} is due to the fact that the l.h.s.\ involves the ZF S-matrix elements related to the permutation of two Fermions. 
The right-hand side is expressed in terms of  the functions
\begin{equation}
\begin{aligned}
    f(x_1^{\pm },x_2^{\pm })=\frac{\sqrt{x_2^{- }}}{\sqrt{x_2^{+ }}} \frac{x_{ 1}^{+}-x_{ 2}^{+}}{x_{1}^{+}-x_{2}^{-}}\,,\qquad
    &&g(x_1^{\pm },x_2^{\pm })=\frac{\sqrt{x_2^{- }}}{\sqrt{x_2^{+ }}}  \frac{1- x^{-}_{1}x^{+}_{2}}{1- x^{-}_{1}x^{-}_{2}}\,,\\
    \tilde{f}(\tx_1^{\pm },\tx_2^{\pm })=\frac{\sqrt{\tx_2^{+ }}}{\sqrt{\tx_2^{- }}} \frac{\tx_{ 1}^{-}-\tx_{ 2}^{-}}{\tx_{1}^{-}-\tx_{2}^{+}}\,,\qquad
    &&\tilde{g}(\tx_1^{\pm },\tx_2^{\pm })=\frac{\sqrt{\tx_2^{+ }}}{\sqrt{\tx_2^{- }}} \frac{1- \tx^{+}_{1}\tx^{-}_{2}}{1- \tx^{+}_{1}\tx^{+}_{2}}\,,
\end{aligned}
\end{equation}
where the Zhukovsky variables may be left or right and related to various masses which here are omitted.

\subsection{Poles and bound states}
\label{sec:ws:poles}
Our proposal for the dressing factors relies on the analytic structure of the S-matrix elements, which for this model has not yet been studied in detail. One crucial piece of information is whether we expect, in a given channel, to encounter a bound-state pole. Those poles should appear when the rapidities go as
\begin{equation}
\label{eq:boundstatecond}
    \text{string}:\quad u_1-u_2\to-\frac{i}{h}(m_1+m_2)\,,\qquad    
    \text{mirror}:\quad u_1-u_2\to+\frac{i}{h}(m_1+m_2)\,,
\end{equation}
in either kinematics.
This is necessary to obtain a supersymmetric (``short'') bound-state representation with physical energy in either kinematics~\cite{Arutyunov:2007tc,Frolov:2025uwz}.
When evaluating the S-matrix on the poles, we expect it to become a projector on a short (two-dimensional) representation with mass $(m_1+m_2)$. This requirement, along with the explicit form of the S-matrix from~\cite{Borsato:2015mma}, is sufficient to rule out bound-state poles in the LR and RL scattering matrices, in either kinematics. We will also rule out any double poles at~\eqref{eq:boundstatecond}, in analogy with $\AdSST$.
Finally, we will assume that in the string kinematics we will encounter poles when scattering two bosons related to the same sphere, based on the semiclassical intuition from giant magnons~\cite{Hofman:2006xt}, and in analogy with $\AdSST$.

\paragraph{String kinematics.}
Similarly to what happens for $\AdSST$ and ($\text{AdS}_5\times\text{S}^5$) we expect to encounter a pole when scattering LL or RR bosonic excitations on the same sphere ($m_1=m_2=m$). Semi-classically, this pole is related to the existence of giant magnons. Therefore, we expect simple poles of the type
\begin{equation}
\label{eq:stringpoles}
    S_{\cX\cX}^{mm}(u_1,u_2)\sim\frac{\text{finite}}{x^{+m}_{\L1}-x^{-m}_{\L2}}\,,\qquad S_{\bar{\cX}\bar{\cX}}^{mm}(u_1,u_2)\sim\frac{\text{finite}}{x^{+m}_{\R1}-x^{-m}_{\R2}}\,,
\end{equation}
as $u_1-u_2\to-2im/h$, where $m=\alpha, 1-\alpha$.
We do not see a reason to expect bound-state poles for scattering involving $m_1\neq m_2$ (i.e., different spheres) in the string kinematics, based on semi-classical intuition.
From~\eqref{eq:stringpoles} and generalised physical unitarity, we also expect the same S-matrix elements to have \textit{a simple zero} as $u_1-u_2\to2im/h$.
Using the explicit form of the S-matrix elements from~\cite{Borsato:2015mma}, it is easy to see what happens to the other S-matrix elements in the same block. One can check that on the bound-state condition the S~matrix becomes a projector on a two-dimensional bound-state representation of mass~$2m$. Furthermore, one sees that the $S_{\Upsilon\Upsilon}^{mm}(u_1,u_2)$ and $S_{\bar{\Upsilon}\bar{\Upsilon}}^{mm}(u_1,u_2)$ elements are necessarily regular as $u_1-u_2\to\pm2im/h$.%
\footnote{This is  different from the case of $\AdSST$ where one would find that, if the scattering of e.g.\ two highest-weight states (the two bosons on the sphere) has a simple pole at $u_1-u_2\to-2im/h$, then the scattering of the two lowest-weight states (the two bosons in AdS) must have a pole at $u_1-u_2\to+2im/h$. This difference is due to the fact that for $\AdSSSS$ the representations are smaller, and the bosons of the spheres are unrelated to those in AdS.}

\paragraph{Mirror kinematics.}
In the mirror kinematics, bound-state poles appear as $u_1-u_2\to+2im/h$. We already know that this cannot happen in the LR and RL channels, and if $m_1=m_2$ (in that case, some S-matrix element vanish as $u_1-u_2\to+2im/h$, but there are  no mirror poles).
The only remaining possibility is to encounter mirror poles when $m_1=\alpha$ and $m_2=1-\alpha$, or viceversa. Indeed, the form of these S-matrix blocks is such that they necessarily have poles in the mirror channel (and therefore zeros in the string one) or poles in the string channel (and therefore zeros in the mirror one). As we did not find a justification for the latter, let us assume that the former is true. In this case, we find that there must be a pole in the fermion-fermion processes,
\begin{equation}
\label{eq:mirrorpoles}
    S_{\Upsilon\Upsilon}^{m_1m_2}(u_1,u_2)\sim\frac{\text{finite}}{\tx^{-m_1}_{\L1}-\tx^{+m_2}_{\L2}}\,,\qquad S_{\bar{\Upsilon}\bar{\Upsilon}}^{m_1m_2}(u_1,u_2)\sim\frac{\text{finite}}{\tx^{-m_1}_{\R1}-\tx^{+m_2}_{\R2}}\,,
\end{equation}
as $u_1-u_2\to i/h$, where $m_1=\alpha$, $m_2=1-\alpha$, or viceversa. Again by unitarity, these S-matrix elements have a simple zero when $u_1-u_2\to -i/h$, while $S_{\cX\cX}^{m_1m_2}(u_1,u_2)$ and $S_{\bar{\cX}\bar{\cX}}^{m_1m_2}(u_1,u_2)$ are necessarily regular on either condition in~\eqref{eq:boundstatecond}.

\section{Dressing factors for massive excitations}
\label{sec:proposal}

In this section we propose S-matrix elements which solve the crossing equations in the mirror kinematics for modes of mass $\alpha$ or $1-\alpha$. The corresponding string S-matrix elements are obtained by an analytic continuation as explained in~\cite{Frolov:2025uwz,Frolov:2025tda}.

\subsection{Proposal}

\paragraph{Same-mass scattering.}
Below we list the scattering elements for processes involving highest-weight states. The remaining processes are fixed by symmetry.
\bal\label{eq:SXXmm}
   S^{m m}_{\cX\cX}(u_1,u_2)=&+H^{mm}_{\cX\cX}  (u_1,u_2)\, {{\tx_{\L1}^{+m}}\ov {\tx_{\L1}^{-m}}}\, {{\tx_{\L2}^{-m}}\ov {\tx_{\L2}^{+m}}}\,\left({ \tx^{-m}_{\L1}- \tx^{+m}_{\L2}\ov \tx^{+m}_{\L1}- \tx^{-m}_{\L2}}\right)^2
    \frac{u_1-u_2 + {2im\ov h}}{u_1-u_2 - {2im\ov h}}\,
    \\
    &\quad\times
    \frac{R(\tg^{+m+m}_{\L\L}) R(\tg^{-m-m}_{\L\L})}{R(\tg^{+m-m}_{\L\L}) R(\tg^{-m+m}_{\L\L}) }
    \left(\frac{\Sigma^{\bes}_{\L\L}(\tx^{\pm m}_{\L1}, \tx^{\pm m}_{\L2})}{\Sigma^{\hl}_{\L\L}(\tx^{\pm m}_{\L1}, \tx^{\pm m}_{\L2})}\right)^{-2}\,,
  \\
 S^{mm}_{\bar{\Upsilon}\mathcal{X}} (u_1,u_2)=&+H^{mm}_{\bar{\Upsilon}\mathcal{X}}
(u_1,u_2)\, {\sqrt{\tilde\a_{\L2}^{+m}}\ov \sqrt{\tilde\a_{\L2}^{-m}}}\, {{\tx_{\R1}^{-m}}\ov {\tx_{\R1}^{+m}}}\, {{\tx_{\L2}^{+m}}\ov {\tx_{\L2}^{-m}}}\,\left({1-{\tx^{+m}_{\R1}} \tx^{-m}_{\L2}\ov  1-{ \tx^{-m}_{\R1}}\tx^{+m}_{\L2}}\right)^2
\\
&\quad\times \frac{R(\tg^{-m+m}_{ \R\L}+ i \pi) R(\tg^{+m-m}_{ \R\L}- i \pi)}{R(\tg^{-m-m}_{ \R\L}+ i \pi) R(\tg^{+m+m}_{\R\L}- i \pi)}\left(\frac{\Sigma^{\bes}_{\R\L}(\tx^{\pm m}_{\R1}, \tx^{\pm m}_{\L2})}{\Sigma^{\hl}_{\R\L}(\tx^{\pm m}_{\R1}, \tx^{\pm m}_{\L2})}\right)^{-2} \,,
\eal
and for convenience let us also write the scattering of lowest-weight states, which follows by symmetry:
\bal\label{eq:SYYmm}
      S^{m m}_{\Upsilon\Upsilon}(u_1,u_2)=&-H^{mm}_{\cX\cX} (u_1,u_2)\, {\sqrt{\tx_{\L1}^{+m}}\ov \sqrt{\tx_{\L1}^{-m}}}\, {\sqrt{\tx_{\L2}^{-m}}\ov \sqrt{\tx_{\L2}^{+m}}}\, { \tx^{-m}_{\L1}- \tx^{+m}_{\L2}\ov \tx^{+m}_{\L1}- \tx^{-m}_{\L2}}
    \frac{u_1-u_2 + {2im\ov h}}{u_1-u_2 - {2im\ov h}}\,
    \\
    &\quad\times
    \frac{R(\tg^{+m+m}_{\L\L}) R(\tg^{-m-m}_{\L\L})}{R(\tg^{+m-m}_{\L\L}) R(\tg^{-m+m}_{\L\L}) }
     \left(\frac{\Sigma^{\bes}_{\L\L}(\tx^{\pm m}_{\L1}, \tx^{\pm m}_{\L2})}{\Sigma^{\hl}_{\L\L}(\tx^{\pm m}_{\L1}, \tx^{\pm m}_{\L2})}\right)^{-2}\,,
    \\
S^{mm}_{\bar\cX\Upsilon}(u_1,u_2)=&+H^{mm}_{\bar{\Upsilon}\mathcal{X}}(u_1,u_2)\,{\sqrt{\tilde\a_{\L2}^{-m}}\ov \sqrt{\tilde\a_{\L2}^{+m}}}\,
{\sqrt{\tx_{\R1}^{+m}}\ov \sqrt{\tx_{\R1}^{-m}}}\, {\sqrt{\tx_{\L2}^{+m}}\ov \sqrt{\tx_{\L2}^{-m}}} \,{{1\ov \tx^{+m}_{\R1}}- \tx^{-m}_{\L2}\ov  {1\ov\tx^{-m}_{\R1}} - \tx^{+m}_{\L2}}\, 
\\
&\quad\times\frac{R(\tg^{-m+m}_{ \R\L}- i \pi) R(\tg^{+m-m}_{ \R\L}+ i \pi)}{R(\tg^{-m-m}_{ \R\L}- i \pi) R(\tg^{+m+m}_{\R\L}+ i \pi)}  \left(\frac{\Sigma^{\bes}_{\R\L}(\tx^{\pm m}_{\R1}, \tx^{\pm m}_{\L2})}{\Sigma^{\hl}_{\R\L}(\tx^{\pm m}_{\R1}, \tx^{\pm m}_{\L2})}\right)^{-2} \,.
\eal
Here $m=\a$ or $m=1-\a$.
The functions $\Sigma^{\bes}$ and $\Sigma^{\hl}$ are modifications of the Beisert-Eden-Staudacher (BES) phase~\cite{Beisert:2006ez} and of the
Hern\'andez-L\'opez (HL)~\cite{Hernandez:2006tk} dressing factors, respectively; they are defined in appendix~\ref{app:def} along with the functions $\tilde{\alpha}^{\pm m}_{\L/\R}$ and $R(\tg)$. Note that our definition of the BES dressing factor differs from the one used for mixed-flux $\AdSST$ in~\cite{Frolov:2025uwz,Frolov:2025tda} by a rescaling of $h$ by~$m$, as detailed in~\eqref{eq:BESm}. 
The functions $H^{mm}$ satisfy the homogeneous crossing equations and the braiding unitarity
\bal
H^{mm}_{\cX\cX} (\bar u_1,u_2)H^{mm}_{\bar{\Upsilon}\mathcal{X}}(u_1,u_2) &=1\,,\qquad H^{mm}_{\cX\cX} ( u_1,u_2)H^{mm}_{\bar{\Upsilon}\mathcal{X}}(\bar u_1,u_2) =1\,,
\\
H^{mm}_{\bar{\Upsilon}\mathcal{X}}&(u_1,u_2) H^{mm}_{{\Upsilon}\bar{\mathcal{X}}}(u_2,u_1)=1\,, 
\eal
where the last equation follows from the braiding unitarity $S^{mm}_{\bar{\Upsilon}\mathcal{X}} (u_1,u_2) S^{mm}_{\mathcal{X}\bar{\Upsilon}}(u_2,u_1)=1$. We will fix them  by comparing the S-matrix elements with perturbative results at the end of the section.

\paragraph{Different-mass scattering.}
In this case we define the S-matrix elements without the BES or HL phase.%
\footnote{Indeed, it is not obvious to us how to generalise those functions to the case of $m_1\neq m_2$; we will return to this point in the conclusions.}
It is quite remarkable that a solution without the BES phase can be found, and that it satisfies all necessary physical requirements. It is
\bal
\label{eq:SXXm1m2}
&S^{m_1m_2}_{\cX\cX} (u_1,u_2)=H^{m_1m_2}_{\cX\cX} (u_1,u_2)\,\frac{R(\tg^{+m_1+m_2}_{\L\L}) R(\tg^{-m_1-m_2}_{\L\L})}{R(\tg^{+m_1-m_2}_{\L\L}) R(\tg^{-m_1+m_2}_{\L\L}) }\,,
\\
& S^{m_1m_2}_{\bar{\Upsilon}\mathcal{X}} (u_1,u_2)=H^{m_1m_2}_{\bar{\Upsilon}\mathcal{X}}
(u_1,u_2)\, {\sqrt{\tilde\a_{\L2}^{+m_2}}\ov \sqrt{\tilde\a_{\L2}^{-m_2}}}\,\frac{R(\tg^{-m_1+m_2}_{ \R\L}+ i \pi) R(\tg^{+m_1-m_2}_{ \R\L}- i \pi)}{R(\tg^{-m_1-m_2}_{ \R\L}+ i \pi) R(\tg^{+m_1+m_2}_{\R\L}- i \pi)} \,,
\eal
\bal
\label{eq:SYYm1m2}
S^{m_1m_2}_{{\Upsilon}{\Upsilon}} (u_1,u_2)=&-H^{m_1m_2}_{\cX\cX} (u_1,u_2)\,{\sqrt{\tx_{\L1}^{-m_1}}\ov \sqrt{\tx_{\L1}^{+m_1}}}\, {\sqrt{\tx_{\L2}^{+m_2}}\ov \sqrt{\tx_{\L2}^{-m_2}}}\,{ \tx^{+m_1}_{\L1}- \tx^{-m_2}_{\L2}\ov \tx^{-m_1}_{\L1}- \tx^{+m_2}_{\L2}}\, 
\\
&\times\frac{R(\tg^{+m_1+m_2}_{\L\L}) R(\tg^{-m_1-m_2}_{\L\L})}{R(\tg^{+m_1-m_2}_{\L\L}) R(\tg^{-m_1+m_2}_{\L\L}) }\,,\\
S^{m_1m_2}_{\bar\cX\Upsilon}(u_1,u_2)=&H^{m_1m_2}_{\bar{\Upsilon}\mathcal{X}}
(u_1,u_2)\, {\sqrt{\tilde\a_{\L2}^{-m_2}}\ov \sqrt{\tilde\a_{\L2}^{+m_2}}}\,{\sqrt{\tx_{\R1}^{-m_1}}\ov \sqrt{\tx_{\R1}^{+m_1}}}\, {\sqrt{\tx_{\L2}^{-m_2}}\ov \sqrt{\tx_{\L2}^{+m_2}}}\,{ {1\ov\tx^{-m_1}_{\R1}} - \tx^{+m_2}_{\L2}\ov {1\ov \tx^{+m_1}_{\R1}}- \tx^{-m_2}_{\L2}}\, 
\\
&\times\frac{R(\tg^{-m_1+m_2}_{ \R\L}- i \pi) R(\tg^{+m_1-m_2}_{ \R\L}+ i \pi)}{R(\tg^{-m_1-m_2}_{ \R\L}- i \pi) R(\tg^{+m_1+m_2}_{\R\L}+ i \pi)}  \,,
\eal
and similar formulae with the replacement $\cX\leftrightarrow \bar\cX$, $\Upsilon\leftrightarrow \bar\Upsilon$. The $H^{m_1m_2}_{ab}$  satisfy the following equations
\bal
H^{m_1m_2}_{\cX\cX} (\bar u_1,u_2)H^{m_1m_2}_{\bar{\Upsilon}\mathcal{X}}(u_1,u_2) &=1\,,\qquad H^{m_1m_2}_{\cX\cX} ( u_1,u_2)H^{m_1m_2}_{\bar{\Upsilon}\mathcal{X}}(\bar u_1,u_2) =1\,,
\\
H^{m_1m_2}_{\bar{\Upsilon}\mathcal{X}}&(u_1,u_2) H^{m_2m_1}_{{\Upsilon}\bar{\mathcal{X}}}(u_2,u_1)=1\,, 
\eal
where the last equation follows from the braiding unitarity $S^{m_1m_2}_{\bar{\Upsilon}\mathcal{X}} (u_1,u_2) S^{m_2m_1}_{\mathcal{X}\bar{\Upsilon}}(u_2,u_1)=1$.

\paragraph{The homogeneous factors.}
The homogeneous factors $H^{mm}$ and $H^{m_1m_2}$ can only be fixed by comparison with perturbative data. As usual in AdS$_3$/CFT$_2$, this means worldsheet computations at strong tension because the weakly-coupled duals are largely unknown or intractable~\cite{Seibold:2024qkh}. Moreover, for $\AdSSSS$ there are not many available  strong-tension perturbative results. We are not aware of any mixed-flux near-BMN computation for this background. The most complete data can be found in~\cite{Bianchi:2014rfa} (see also references therein). Expanding our proposal as explained in appendix~\ref{app:BMN} we find a simple solution for the homogeneous dressing factors:
\begin{equation}\label{eq:Hm1m2}
\begin{aligned}
    &H^{mm}_{\cX\cX}  (u_1,u_2)\!\!\!\!&&=H^{mm}_{\bar{\Upsilon}\cX}  (u_1,u_2) \!\!\!\!&&= e^{  -\frac{i}{2}\frac{1-m}{m}(\tp_1\tE_2 -\tp_2\tE_1)} \,,\\
    &H^{m_1m_2}_{\cX\cX}  (u_1,u_2)\!\!\!\!&&=H^{m_1m_2}_{\bar{\Upsilon}\cX}  (u_1,u_2) \!\!\!\!&&= e^{  \frac{i}{2}(\tp_1\tE_2 -\tp_2\tE_1)} \,,\qquad m_1\neq m_2\,,
\end{aligned}
\end{equation}
valid for any value $\a,1-\a$ of the masses, where $\tE$ is the mirror energy in the left or right kinematics depending on the particle.

\paragraph{String kinematics.}
To obtain the dressing factors in the string kinematics is is necessary to perform an analytic continuation following Section 4.5 of~\cite{Frolov:2025uwz}. For the factors $H^{m_1m_2}$, which are written in terms of (mirror) energy and momentum, the continuation to the string region follows from
\begin{equation}\label{eq:Hm1m2string}
    \tp_1\tE_2 -\tp_2\tE_1\to p_1 H_2 -p_2 H_1\,.
\end{equation}
Similarly, for any meromorphic function of $\tilde{x}^{\pm m}$ and $\tilde{\gamma}^{\pm m}$ we can simply perform the replacement
\begin{equation}
    \tilde{x}^{\pm m}\to x^{\pm m}\,,\qquad
    \tilde{\gamma}^{\pm m}\to \gamma^{\pm m}\,,
\end{equation}
while the analytic continuation of the BES and HL phases requires more care. This continuation is worked out in detail in appendix G of~\cite{Frolov:2025uwz}, where in particular the result for the S-matrix element of type $\cX\cX$ is given, cf.~(G.15) there.

\paragraph{Interpretation of the CDD factors.}
The CCD factors $H^{mm}$ and $H^{m_1m_2}$ are quite unusual when comparing with $\AdSST$, and deserve further discussion. The antisymmetric combination $\tp\wedge\tE$, or in string theory  $p\wedge H$, is precisely the type of CDD factor that we would expect from a $T\overline{T}$ deformation~\cite{Smirnov:2016lqw,Cavaglia:2016oda}. If we were considering a ``standard'' (relativistic and local) integrable QFT, a $T\overline{T}$ deformation would spoil the large-rapidity (or large-energy) behaviour of the model. Indeed, for a local relativistic QFT we would expect $ S(\theta_{12})$ to have a finite limit as $\theta_{12}\to\pm\infty$,%
\footnote{Interestingly, the two limits need not coincide and need not be the identity~\cite{Frolov:2025ozz}.}
something which is not preserved by the $T\overline{T}$ CDD factor which has an essential singularity in either limit.
However, the QFT that lives on the worldsheet of the string in lightcone gauge is not a local theory. This is particularly transparent in the case of flat space~\cite{Dubovsky:2012wk}. More generally, different choices of the uniform lightcone gauge~\cite{Arutyunov:2005hd} would change the CDD factor precisely by a term proportional to $p\wedge H$~\cite{Baggio:2018gct}. 
It is however interesting to note that for $\AdSST$ (and for $\text{AdS}_5\times\text{S}^5$) there exists \textit{one particular choice of uniform lightcone gauge} so that the two-particle S~matrix has a good large-energy limit. This can only be seen in the mirror kinematics (because in the string kinematics, at least for RR backgrounds, both $p$ and $H$ are bounded) by taking the limit $\tp_1\to+\infty$ and $\tp_2\to-\infty$. Curiously or perhaps suspiciously, the same is not possible for $\AdSSSS$ precisely due to the factors $H^{mm}$ and $H^{m_1m_2}$.
Nevertheless, we do not see a clear reason \textit{a priori} to exclude such a factor. Moreover, as we will see below, these factors ensure the correct behaviour with respect to symmetries.

\paragraph{Symmetry descendants.}
It is also interesting to consider the behaviour of the S-matrix elements when one of the momenta is zero. We expect this to be related to the action of the symmetries of $\mathfrak{d}(2,1;\alpha)^{\oplus2}$. To see this, imagine that we have a state $|\Psi_N\rangle$ identified by a given collection of $N$~excitations whose momenta and rapidities (for auxiliary particles) solve the Bethe--Yang equations.%
\footnote{In principle, we should consider the excited-state mirror TBA equations to describe how symmetries act on non-perturbative states~\cite{Arutyunov:2011uz}, but we would expect that the symmetry structure should be apparent from the Bethe--Yang equations, like in $\text{AdS}_5\times\text{S}^5$~\cite{Beisert:2005fw}.}
We can easily read off
\begin{equation}
\label{eq:chargesBetheState}
\begin{aligned}
    \left(\mathbf{L}_0-\alpha\,\mathbf{J}^{3}_{(1)}-(1-\alpha)\,\mathbf{J}^{3}_{(2)}\right)|\Psi_{N}\rangle&=\frac{1}{2}\left(H_{\text{tot}}+M_{\text{tot}}\right)|\Psi_{N}\rangle\,,\\
    \left(\tilde{\mathbf{L}}_0-\alpha\,\tilde{\mathbf{J}}^{3}_{(1)}-(1-\alpha)\,\tilde{\mathbf{J}}^{3}_{(2)}\right)|\Psi_{N}\rangle&=\frac{1}{2}\left(H_{\text{tot}}-M_{\text{tot}}\right)|\Psi_{N}\rangle\,,
\end{aligned}
\end{equation}
where $\mathbf{L}_0,\tilde{\mathbf{L}}_0$ are the left and right $\mathfrak{sl}(2,\mathbb{R})$ Cartan elements, respectively, and $\mathbf{J}^3_{(j)},\tilde{\mathbf{J}}^3_{(j)}$ are the Cartan elements for the $j$-the sphere. The values of $H_{\text{tot}}$ and $M_{\text{tot}}$ can be found from~\eqref{eq:representation-charges} by summing over all particles' momenta. Suppose that we want to create a symmetry descendant of this state by acting with the lowering operator $\mathbf{J}^-_{(1)}$ or $\mathbf{J}^-_{(2)}$, respectively, and that this does not yield a null state. The second line of \eqref{eq:chargesBetheState} must be unchanged, while the first should change precisely by $\alpha$ or $(1-\alpha)$, respectively. By looking at eq.~\eqref{eq:representation-charges} we see that this can be achieved by adding to the state one bosonic excitation with $\mu=\alpha,(1-\alpha)$, respectively, and $p=0$. In a similar way, it is easy to convince oneself that a descendant of $\tilde{\mathbf{J}}^-_{(1)}$ or $\tilde{\mathbf{J}}^-_{(2)}$ should be obtained by adding one bosonic excitation $\mu=-\alpha,-(1-\alpha)$, respectively, and $p=0$. The case of supersymmetry descendants can also be treated similarly, but it requires adding also an auxiliary root along with the massive excitations; we will not present this here as it is not crucial to the issue which we want to highlight.
The crucial question is: \textit{does the new state with $(N+\delta N)$ excitations, $N$ of which have the same momenta/rapidities as $|\Psi_N\rangle$, and $\delta N$ of which have $p=0$, automatically solve the Bethe--Yang equations?} This must be the case to have a descendant.
Let us illustrate this on the example of a state containing $N_Y$ particles of type $Y$ and $N_X$ particles of type $X$, and for simplicity nothing else ($N=N_X+N_Y$). Originally, the Bethe equations have the form
\begin{equation}
\label{eq:originalBethe}
\begin{aligned}
    &1=e^{ip_aL}\prod_{b=1}^{N_Y}S_{YY}(p_a,p_b)\prod_{b=1}^{N_X}S_{YX}(p_a,p_b)\,,\qquad a=1,\dots, N_Y\,,\\
    &1=e^{ip_cL}\prod_{b=1}^{N_Y}S_{XY}(p_c,p_b)\prod_{b=1}^{N_X}S_{XX}(p_c,p_b)\,,\qquad c=1,\dots, N_X\,,
\end{aligned}
\end{equation}
and we assume that they are satisfied. Consider the state with $\delta N_Y$ additional particles of type $Y$ with $p=0$ and $\delta N_X$ additional particles of type $X$ with $p=0$. This should correspond to acting on the original state with $(\mathbf{J}^{-}_{(1)})^{\delta N_Y}(\mathbf{J}^{-}_{(2)})^{\delta N_X}$. The new Bethe equations are
\begin{equation}
\label{eq:newBethe}
\begin{aligned}
    &1=e^{ip_aL'}S_{YY}(p_a,0)^{\delta N_Y}S_{YY}(p_a,0)^{\delta N_X}\prod_{b=1}^{N_Y}S_{YY}(p_a,p_b)\prod_{b=1}^{N_X}S_{YX}(p_a,p_b)\,,\quad a=1,\dots N_Y\,,\\
    &1=e^{ip_cL'}S_{XY}(p_c,0)^{\delta N_Y}S_{XX}(p_c,0)^{\delta N_X}\prod_{b=1}^{N_Y}S_{XY}(p_c,p_b)\prod_{b=1}^{N_X}S_{XX}(p_c,p_b)\,,\qquad c=1,\dots, N_X\,,
\end{aligned}
\end{equation}
supplemented by $(\delta N_Y+\delta N_X)$ equations for the zero-modes themselves
\begin{equation}
\label{eq:descendent1}
\begin{aligned}
    &1=\prod_{b=1}^{N_Y+\delta N_Y}S_{YY}(0,p_b)\prod_{b=1}^{N_X+\delta N_X}S_{YX}(0,p_b)\,,\\
    &1=\prod_{b=1}^{N_Y+\delta N_Y}S_{XY}(0,p_b)\prod_{b=1}^{N_X+\delta N_X}S_{XX}(0,p_b)\,.
\end{aligned}
\end{equation}
Notice that because the length $L$ is the eigenvalue of the charge $(\alpha\,\mathbf{J}^{3}_{(1)}+(1-\alpha)\,\mathbf{J}^{3}_{(2)})$ in the lightcone gauge, we have that on the new state
\begin{equation}
    L'=L-\alpha\,\delta N_Y-(1-\alpha)\,\delta N_X.
\end{equation}
This observation, and the original Bethe equations~\eqref{eq:originalBethe}, can be used to recast~\eqref{eq:newBethe} as
\begin{equation}
\label{eq:descendent2}
\begin{aligned}
&1=\Big[e^{-i\alpha p}S_{YY}(p,0)\Big]^{\delta N_Y}
\Big[e^{-i(1-\alpha) p}S_{YX}(p,0)\Big]^{\delta N_X},\\
&1=\Big[e^{-i\alpha p}S_{XY}(p,0)\Big]^{\delta N_Y}\Big[e^{-i(1-\alpha) p}S_{XX}(p,0)\Big]^{\delta N_X},
\end{aligned}
\end{equation}
valid for any~$p$. This is a constraint on the dressing factors. Expanding our proposed dressing factors we find
\begin{equation}
\label{eq:Spzero}
\begin{aligned}
    S_{YY}(p,0)&=e^{\frac{i}{2}(1+\alpha)p},
    \qquad&S_{YX}(p,0)&=e^{\frac{i}{2}(1-\alpha)\,p},\\
    S_{XY}(p,0)&=e^{\frac{i}{2}\alpha\,p},
    \qquad&S_{XX}(p,0)&=e^{\frac{i}{2}(2-\alpha)p}.
\end{aligned}
\end{equation}
We see that equations~\eqref{eq:descendent2} are only solved if $\delta N_X=\delta N_Y$, i.e.\ if we change the angular momenta of the two spheres at the same time. Moreover, equations~\eqref{eq:descendent1} require the original state to satisfy a condition stronger than level-matching: the sum of all momenta of $X$-particles and that of momenta of $Y$-particle must vanish \textit{separately},
\begin{equation}
\label{eq:stronglvlmatch}
    \sum_{a=1}^{N_Y}p_a=\sum_{a=1}^{N_X}p_a=0\,.
\end{equation}
This is puzzling, and means broken symmetries are not simply restored by adding zero-momentum particles, like we would have expected following e.g.~\cite{Beisert:2005fw}. This puzzle is not unique for our proposal: rather, we inherit it from the tree-level results of~\cite{Rughoonauth:2012qd}. It is easy to see that eqs.~\eqref{eq:Spzero} match with the tree level results,%
\footnote{In fact, taking the near-BMN limit and setting one of the momenta to zero are commuting operations, at least in the pure-RR case.} so the same issue is visible already at tree level (and was indeed discussed in~\cite{Rughoonauth:2012qd}). One possible explanation is that in the computation of the S-matrix (perturbative or exact), one takes the decompactification limit by taking the charge for each of the four generators $\mathbf{J}_{(j)}^{3}$ and  $\tilde{\mathbf{J}}_{(j)}^{3}$ to be equal to $L/2$ and then sending $L\to\infty$. Aside from the combination of R-charges in the definition of the length~$\mathbf{L}$, there is an orthogonal combination~$\mathbf{N}$
\begin{equation}
\mathbf{L}=\alpha\mathbf{J}_{(1)}^{3}+(1-\alpha)\,\mathbf{J}_{(2)}^{3}+\alpha\tilde{\mathbf{J}}_{(1)}^{3}+(1-\alpha)\tilde{\mathbf{J}}_{(2)}^{3}\,,\qquad
    \mathbf{N}=\mathbf{J}_{(1)}^{3}-\mathbf{J}_{(2)}^{3}+\tilde{\mathbf{J}}_{(1)}^{3}-\tilde{\mathbf{J}}_{(2)}^{3}\,.
\end{equation}
The massless modes $T,\bar{T}$ are themselves linear combinations of the coordinate on $\text{S}^1$ and of the coordinate which has $\mathbf{N}$ as isometry. Hence, changing the eigenvalue of $\mathbf{N}$ requires \textit{changing the momentum of the massless modes} $T,\bar{T}$. Unfortunately, it is not clear how to encode the momentum of a massless field in the Bethe--Yang (or TBA) equations, even in the simpler case of $\AdSST$.
This might explain why only when adding particles in pairs from either sphere ($\delta N_X=\delta N_Y$), without changing the eigenvalue of $\mathbf{N}$, we find a straightforward solution of eq.~\eqref{eq:descendent1}. However, this does not yet explain why we (as well as~\cite{Rughoonauth:2012qd}) encounter the condition~\eqref{eq:stronglvlmatch}. What is its origin in string theory?
Taken at face value, this discussion indicates that \textit{we do not know how the $\mathfrak{d}(2,1;\alpha)$ symmetry (or even its bosonic piece) is restored in the string spectrum from the lightcone gauge fixing.}%
\footnote{%
This issue might also underlie the puzzles that emerge in the QSC description~\cite{Cavaglia:2025icd,Chernikov:2025jko} which we will summarise in section~\ref{sec:conclusions}.}
Because the issue is already manifest in perturbative computations, we are hopeful that it may be solved by a careful analysis of the  quantisation of the near-BMN expansion at tree level.%
\footnote{It is worth noting that even at tree level this analysis is non-trivial precisely due to the zero-momentum massless modes, that appear in cubic interactions. Their study requires a regularisation. The approach used to compute the S-matrix in~\cite{Rughoonauth:2012qd} and in subsequent works is natural for integrability, as it uses a regularisation which is compatible with the classical spectral curve.}

\subsection{Bound states and Bethe strings}
In accordance with our discussion of section~\ref{sec:ws:poles}, we have string-kinematics poles in $S_{\cX\cX}^{mm}$, which are quite reminiscent of what happens for $\AdSST$. It is easy to see that such poles can be used to construct bound states of mass $2m,3m, \dots$, which sit in a two-dimensional representation whose highest-weight state is $\cX\otimes\cdots\otimes\cX$.

The study of bound states and Bethe strings in the mirror kinematics is more interesting because it is what yields the mirror TBA equations.
Since $S^{m_1m_2}_{{\Upsilon}{\Upsilon}}$ (with $m_1\neq m_2$) has a mirror bound-state pole at $\tx^{-m_1}_{\L1}=\tx^{+ m_2}_{\L2}$,  we might expect to find a bound-state representation with lowest-weight state $\cZ(u)\sim \Upsilon^{(m_1)}(u_1)\Upsilon^{(m_2)}(u_2)$ of mass $m_1+m_2=1$. However, it can be checked that this configuration does not generically solve the mirror Bethe-Yang equations with an arbitrary number of other particles. To understand the issue, consider the fused S-matrix of a ``$\cZ$-particle'' with, say, a $\Upsilon$-particle of mass $m_1$
\bal
\label{eq:wrongfusion}
S^{1m_1}_{\cZ\Upsilon}(u,v)= S^{m_1m_1}_{\Upsilon\Upsilon}(u_1,v)\,S^{m_2m_1}_{\Upsilon\Upsilon}(u_2,v)\,,
\eal
where, owing to the bound-state condition~\eqref{eq:boundstatecond}
\bal
u_1 = u+{i\ov h} m_2\,,\qquad u_2 = u-{i\ov h}m_1\,.
\eal
It is straightforward to check that such a ``fused S-matrix element'' does not satisfy physical unitarity. This is the reason why this configuration does not solve the Bethe-Yang equations.

This situation is reminiscent of what happens in $\text{AdS}_4\times \mathbb{C}\text{P}^3$, and the correct Bethe strings actually involve \textit{two more excitations}, replacing~\eqref{eq:wrongfusion} with
\begin{equation}
\label{eq:rightfusion}
    S^{1m_1}_{\cZ\Upsilon}(u,v)S^{1m_1}_{\cZ'\Upsilon}(u,v)= S^{m_1m_1}_{\Upsilon\Upsilon}(u_1,v)S^{m_2m_1}_{\Upsilon\Upsilon}(u_2,v)S^{m_2m_1}_{\Upsilon\Upsilon}(u_1',v)S^{m_1m_1}_{\Upsilon\Upsilon}(u_2',v)\,,
\end{equation}
where
\begin{equation}
    u_1 = u+{i\ov h} m_2\,,\qquad u_2 = u-{i\ov h}m_1\,,\qquad
    u_1' = u+{i\ov h} m_1\,,\qquad u_2' = u-{i\ov h}m_2\,,
\end{equation}
so that the product in~\eqref{eq:rightfusion} is unitary. In~\cite{Bombardelli:2009xz}, similar configurations in $\text{AdS}_4\times \mathbb{C}\text{P}^3$ were also observed, and dubbed ``strange Bethe strings''. Borrowing the language of that paper, we list below the  Bethe strings appearing in the mirror kinematics.

\paragraph{Strange strings.}
We can generalise the case discussed above to a configuration of $4Q$ particles, obtained by putting together this $2Q$-particle complex
\begin{equation}
\begin{aligned}
\text{type }m_1:\quad
\tx_j^{\pm m_1}=\tx(u_j\pm \tfrac{i}{h}m_1)\,,&\qquad u_j =u+\frac{i}{h}(Q+2-m_1-2j)\,,\quad &&j=1,\dots Q\,,\\
\text{type }m_2:\quad
\tx_{j'}^{\pm m_2}=\tx(u_{j'}\pm \tfrac{i}{h}m_2)\,,&\qquad u_{j'} =u+\frac{i}{h}(Q+m_2-2j')\,,\quad &&j'=1,\dots Q\,.
\end{aligned}
\end{equation}
with its complex conjugate one:
\begin{equation}
\begin{aligned}
\text{type }m_1:\quad
\tx_j^{\pm m_1}=\tx(u_j\pm \tfrac{i}{h}m_1)\,,&\qquad u_j =u+\frac{i}{h}(Q+m_1-2j)\,,\quad &&j=1,\dots Q\,,\\
\text{type }m_2:\quad
\tx_{j'}^{\pm m_2}=\tx(u_{j'}\pm \tfrac{i}{h}m_2)\,,&\qquad u_{j'} =u+\frac{i}{h}(Q+2-m_2-2j')\,,\quad &&j'=1,\dots Q\,.
\end{aligned}
\end{equation}

\paragraph{Wide strings with $m_1$-type centre.}
This configuration is symmetric about the real line, and it consists of excitations
\begin{equation}
\begin{aligned}
\text{type }m_1:\quad
\tx_j^{\pm m_1}=\tx(u_j\pm \tfrac{i}{h}m_1)\,,&\qquad u_j =u+\frac{i}{h}(Q+1-2j)\,,\quad &&j=1,\dots Q\,,\\
\text{type }m_2:\quad
\tx_{j'}^{\pm m_2}=\tx(u_{j'}\pm \tfrac{i}{h}m_2)\,,&\qquad u_{j'} =u+\frac{i}{h}(Q-2j')\,,\quad &&j'=1,\dots Q-1\,.
\end{aligned}
\end{equation}

\paragraph{Wide strings with $m_2$-type centre.}
This configuration is symmetric about the real line, and it is obtained by swapping the particles with respect to the ones above:
\begin{equation}
\begin{aligned}
\text{type }m_1:\quad
\tx_j^{\pm m_1}=\tx(u_j\pm \tfrac{i}{h}m_1)\,,&\qquad u_j =u+\frac{i}{h}(Q-2j)\,,\qquad &&j=1,\dots Q-1\,,\\
\text{type }m_2:\quad
\tx_{j'}^{\pm m_2}=\tx(u_{j'}\pm \tfrac{i}{h}m_2)\,,&\qquad u_{j'} =u+\frac{i}{h}(Q+1-2j')\,,\qquad &&j'=1,\dots Q\,.
\end{aligned}
\end{equation}

\paragraph{Particles.}
On top of these Bethe strings, the mirror TBA will feature fundamental particles (left and right) of mass $m=\a, 1-\a,0$ as well as auxiliary particles.

\paragraph{Fusion.} 
It is easy (but cumbersome) to work out the fused S-matrix elements from these formulae above. It is however interesting to note that --- unlike what happens in other instances of integrable AdS/CFT --- the fused S-matrix elements \textit{still depend on the Bethe-string state constituents}. Ideally, one might have expected the products of the various factors to (largely) telescope, so that the final results depend only on the rapidity~$u_j$ and Bethe-string size $Q_j$. It is not clear to us if this is a fundamental issue of this proposal, but it certainly makes the derivation of the mirror TBA much more cumbersome.
Because the fused S-matrix elements depend on the bound-state constituents, one may worry that some of the expressions may be ambiguous. In particular,
in the mirror theory and for $\alpha=1/2$ one might encounter e.g.~$\tx^{-m_1}_{\L}(u+im_2/h)=\tx_{\L}(u)$ in expressions of the form~\eqref{eq:rightfusion}. When $u$ is taken to be real, as required by physical unitarity,  $\tx_{\L}(u)$ is evaluated on its cut. This would create an ambiguity. However upon closer inspection one finds that all expressions involving $\tx_{\L}(u)$ cancel out in the fused result. This is an encouraging property of our proposal.


\section{Conclusions and outlook}
\label{sec:conclusions}

We have proposed a solution of the crossing equations for strings on mixed-flux $\AdSSSS$ backgrounds valid for any $0<\alpha<1$. Our proposal focuses on the massive excitations, but it should be possible to extend it to massless modes too similarly to what was done in $\AdSST$~\cite{Frolov:2025tda}. It is worth noting that in~$\AdSST$ there is evidence that massless particles obey a non-trivial braid statistics~\cite{Frolov:2025ozz}, which has implications for crossing equations and their solutions. It would be important to understand if this can be the case here.

One might wonder whether, as a check of our construction, we should demand that the phases of $\AdSSSS$ are related to those of $\AdSST$ as $\alpha\to0$ (or $\alpha\to1$). Recall that in those limits, either sphere blows up, resulting in a geometry of the type $\text{AdS}_3\times\text{S}^3\times\mathbb{R}^3\times\text{S}^1$, which in practice is often treated like $\AdSST$ when restricting to the perturbative string spectrum with no momentum or winding on the torus.
It is certainly true that the $\alpha\to0$ and $\alpha\to1$ limits should match with the $\AdSST$ results \textit{in the near-BMN limit}. This is because the near-BMN Lagrangian is regular in $\alpha$, and the limit can be taken without issue. This check is implicit in our check of the near-BMN expansion in appendix~\ref{app:BMN}. It is the case, however, that the all-loop $\AdSSSS$ S~matrix cannot go to the $\AdSST$ one as $\alpha\to0$ or $\alpha\to1$. The structure of the representations (and even the notion of which particles are physical and which ones are not) changes drastically at $\alpha=0$ and so do the crossing equations.
To see this, one may consider the double-crossing equation for, e.g., the $\AdSSSS$ S-matrix element $S^{\a,\a}_{YY}(u_1,u_2)$ and compare them with those of the $\AdSST$ S-matrix element $S_{YY}(u_1,u_2)$ (which na\"ively would be its ``limit'' as $\alpha\to1$). It is straightforward to check that the \textit{two double crossing equations are different} (the latter is the square of the former), which shows that the S-matrix elements cannot be equal --- even in this limit.

One feature of the near-BMN expansion which deserves further investigation is how the $\mathfrak{d}(2,1;\alpha)^{\oplus2}$ symmetry may be restored in the spectrum. As we discussed here (and as already noted in~\cite{Rughoonauth:2012qd}), not all symmetry descendants can be realised by adding zero-momentum excitations. It would be interesting to carry out a detailed analysis of the broken $\mathfrak{d}(2,1;\alpha)^{\oplus2}$ generators in the lightcone gauge to understand their action on the vacuum and on excited states, at least at tree level.

When proposing a solution to the crossing equations, a natural question is whether under appropriate assumptions it is unique. For AdS/CFT worldsheet models this is very hard to prove due to the very involved analytic structure of the dressing factors. In the case at hand, we observe that our solution appears to be unique if we rule out a dressing factor of the BES type between particles related to different spheres.  This seems reasonable because the appearance of the AFS phase~\cite{Arutyunov:2004vx} is tied to semiclassical solutions which live in one sphere, rather than across two spheres, and because the  BES phase~\cite{Beisert:2006ez} is the natural all-loop extension of the AFS one.
Nonetheless, other solutions of the crossing equations which are compatible with the near-BMN perturbative expansion may be found. For general $\alpha$, these would require generalising the BES phase to the scattering of different masses. It is known how to do this for the scattering of bound states in (mixed-flux)~$\AdSST$~\cite{Frolov:2025uwz}. In the mixed-flux case, recall that $\alpha=k_2/(k_1+k_2)$ and $1-\alpha=k_1/(k_1+k_2)$. It is tempting to define a ``fundamental particle'' of minimal mass $m_{\text{min}}=1/(k_1+k_2)$ and treat all particles as if they were its bound states.
It would be then straightforward to define the BES factor for particles of mass~$m_{\text{min}}$ and their bound states. 
Only a fraction of these bound states would be physical, which may require introducing a restricted Hilbert space like in refs.~\cite{Colomo:1991gw,Frolov:2025ozz}.
This is an interesting idea which should be explored in the future. However, to make things simpler and address the question whether other solutions exist, it is more practical to consider the case of $\alpha=1/2$ as we do in appendix~\ref{app:solutions}. We find that other solutions do exist: one involves introducing the BES factor in \textit{all scattering processes} (``same'' or ``different'' spheres)  see appendix~\ref{app:solutions:1}, and the other involves doing so \textit{only for different spheres}, see appendix~\ref{app:solutions:2}. The former sounds  most appealing because one would expect the AFS/BES phase to play a role for the scattering on the same sphere. However, that proposal has a shortcoming as the fused S-matrix elements depend on the bound-state constituents in an ambiguous way.%
\footnote{Some of the Zhukovsky variables and dressing factors are evaluated on their cut, which creates an ambiguity. It is not obvious that one can pick a side of the cut in a way which is unambiguous and compatible with physical unitarity, unlike what happens for our main proposal, see section~\ref{sec:proposal}.}

Two proposals for the quantum spectral curve, that appear to coincide, were recently put forward for pure-RR backgrounds and $\alpha=1/2$~\cite{Cavaglia:2025icd,Chernikov:2025jko}. In principle, these proposals should contain information on the dressing factors, at least for massive excitations. Let us compare those proposals to ours, to the extent that it is possible. When $\alpha=1/2$, both~\cite{Cavaglia:2025icd,Chernikov:2025jko} remark  that the QSC construction does not yield a crossing-symmetric factor which also satisfies braiding unitarity. This is quite unusual, and so far not explained by first principles. By contrast, our solution satisfies both crossing and braiding unitarity (as well as physical unitarity) for any~$\alpha$.
It is not completely straightforward, however,  to extract the dressing factors from the QSC in terms of the BES factor and  explicit special functions like we did here.%
\footnote{We are very grateful to the authors of~\cite{Cavaglia:2025icd} for providing further details and clarification on the form of the dressing factors implied by their QSC construction in private correspondence.} 
As far as we can verify, it appears that the QSC prediction for the dressing factors~\cite{Cavaglia:2025icd} is not compatible with ours even when restricting to~$\alpha=1/2$ (and pure-RR backgrounds). It rather appears that the solution of~\cite{Cavaglia:2025icd} may be related to the one which we present in appendix~\ref{app:solutions:2}. However, the QSC proposal does not satisfy the standard  crossing and braiding unitarity conditions of~\cite{Borsato:2015mma}. We present a more detailed comparison of the solution in appendix~\ref{app:solutions:2} and the QSC proposal in appendix~\ref{app:scomparison}; there, we show that a certain ``symmetric'' combination of the dressing factors \textit{almost} matches in the two proposals, up to a solution of the homogeneous crossing equation which is necessary to include for consistency with near-BMN  perturbative calculations.
Let us also emphasise that $\alpha\neq1/2$, a complete proposal of the QSC is not known yet. Ref.~\cite{Cavaglia:2025icd} proposed the $Q$-system and presented some preliminary work towards the QSC  for general~$\alpha$, but given the issues with unitarity and crossing we do not expect a match of the dressing factors in that case either.

Our own proposal contains some unusual features which should be understood better. First of all, we need to introduce a CDD factor whose phase is proportional to $\tp_1\tE_2-\tp_2\tE_1$. This is the same function which appears in $T\bar{T}$-deformed theories~\cite{Cavaglia:2016oda,Smirnov:2016lqw}, and in the uniform lightcone gauge (not coincidentally~\cite{Baggio:2018gct,Frolov:2019nrr}). This term is compatible with the expected properties of the worldsheet S~matrix, but it usually appears from the asymptotic expansion of the BES phase, rather than being added ``by hand'' as a CDD factor. It would be interesting to understand if the phases can be redefined so that such a factor is incorporated more naturally.
A second unusual feature is that the fusion of the BES factors, and of the expressions which depend on the Zhukovsky variables, does not telescope completely --- the result depends on the constituent particles. This is unlike what happens in other instances of integrable AdS/CFT. Still, this does not create any ambiguity in and of itself. It is possible that a different way to express the dressing factors may cancel the dependence on constituents, similar to the ``improved'' dressing factor of~\cite{Arutyunov:2009kf}, but we were unable to find such a form.

We hope to return to some of these questions in the future.

\section*{Acknowledgments}
We thank the anonymous referee for raising several interesting points in their report, which we addressed in this version of the paper.
We are very grateful to Davide Polvara for collaboration at an early stage of this project and for numerous useful discussions throughout. We thank Andrea Cavagli\`a, Rouven Frassek, Nicol\`o Primi, and Roberto Tateo for correspondence regarding their recent work and on the identification of the dressing factors therein.
We thank the organisers of the \textit{Integrability, Dualities and Deformations 2025} Workshop at NORDITA, Stockholm for hospitality and for the stimulating environment which contributed to this work.
A.S.\ was supported in part by the CARIPARO Foundation Grant under grant n.~68079.
S.F.\ acknowledges support from the INFN under a Foreign Visiting Fellowship in the initial stages of this work.

\appendix
\section{Notation and useful definitions}
\label{app:def}
We collect here some useful formulae, see~\cite{Frolov:2025uwz,Frolov:2025tda} for more details.

\paragraph{String Zhukovsky variables.}
In the string kinematics, the Zhukovski variables can be expressed in terms of momenta as
\begin{equation}
\label{eq:xpmofp_app}
\begin{aligned}
    x^{\pm m}_{\L} (p)&=& \frac{e^{\pm i p /2}}{2h\,\sin \tfrac{p}{2}} \Big(m+\frac{k}{2\pi}p+\sqrt{\big(m+\tfrac{k}{2\pi}p\big)^2+4h^2\sin^2\tfrac{p}{2}}\Bigg)\,,\\
    x^{\pm m}_{\R}(p) &=& \frac{e^{\pm i p /2}}{2h\,\sin \tfrac{p}{2}} \Big(m-\frac{k}{2\pi}p+\sqrt{\big(m-\tfrac{k}{2\pi}p\big)^2+4h^2\sin^2\tfrac{p}{2}}\Bigg)\,.
\end{aligned}
\end{equation}
They obey
\begin{equation}
x^{+m}_{a}+\frac{1}{x^{+m}_{a}}-x^{-m}_{a}-\frac{1}{x^{-m}_{a}}=\frac{2i}{h}m + i\frac{\ka_a}{\pi}p\,,\qquad e^{ip}={x^{+m}_{a}\ov x^{-m}_{a}}\,,
\end{equation}
with $a=$L,R, and the lightcone energy is
\begin{equation}
    H_a(p)=\frac{h}{2i}\left(x^{+m}_{a}-\frac{1}{x^{+m}_{a}}-x^{-m}_{a}+\frac{1}{x^{-m}_{a}}\right)=\sqrt{\big(\mu-\tfrac{k}{2\pi}p\big)^2+4h^2\sin^2\tfrac{p}{2}}\,,
\end{equation}
where $\mu=\pm m$ depending on whether $a=$L,R, respectively.

\paragraph{Rapidity.}
The Zhukovsky variable can be expressed  in terms of the rapidity~$u$
\begin{equation}
x_{a}^{\pm m}(u)=x_{a}(u\pm {i\ov h}m),\qquad
p_a = i\,(\ln x_a^{-m} - \ln x_a^{+m})\,,
\end{equation}
where~\cite{Stepanchuk:2014kza,Frolov:2023lwd}
\begin{equation}
\label{eq:urapidity}
u_a(x)=x+\frac{1}{x}- \frac{\kappa_a}{\pi}\,\ln x \,,
\end{equation}
where $\ka_{\L}=-\ka_{\R}=\frac{k}{h}$ (with $k=1,\,2, \dots$ and $h \ge0$).
The image of the main branch point on the $x$-plane is at
\begin{equation}
\xi_\L =\xi\,,\qquad \xi_\R={1\ov \xi}\,,\qquad  \xi\equiv {\ka\ov2\pi}+ \sqrt{1+{\ka^2\ov4\pi^2}}\,.
\end{equation}

\paragraph{Mirror Zhukovsky variables.}
It is possible to invert~\eqref{eq:urapidity} in a different way, which yields
\begin{equation}
    \tx_a(u)=\begin{cases}
        x_a(u) &\text{Im}[\tx_a]>0\,,\\
        1/x_a(u) &\text{Im}[\tx_a]<0\,,
    \end{cases}
\end{equation}
and it is possible to express the mirror kinematics by means of this variable. The mirror transformation gives
\bal\label{eq:def_mirror_en_mom}
{1\ov i}p_a\ \mapsto\  \tE_a = \ln \tx_a^{-m} - \ln \tx_a^{+m} \,,\qquad
{1\ov i}E_a \ \mapsto\  \tp_a &= {h\ov2}\left( \tx_a^{-m} -{1\ov \tx_a^{-m}} - \tx_a^{+m} +{1\ov \tx_a^{+m}} \right)
\,.
\eal
which satisfies the reality conditions
\bal
\left(\tE_a(u) \right)^* = \tE_{\bar a}(u^*)\,, \quad \left(\tp_a(u) \right)^* = \tp_{\bar a}(u^*)\,.
\eal
In the mirror kinematics, it is not possible to write explicitly $\tilde{x}^\pm_a(\tp)$ as we did in the string kinematics in~\eqref{eq:xpmofp_app}.

\paragraph{Barnes factor and $\gamma$ rapidities.}
In the string kinematics, we introduce $\g$-rapidities by
\bal
x_a(\g)&={\xi_a+e^{\g} \ov 1 - \xi_a e^{\g} }\,,\qquad  
\g_a(x) = \ln{x - \xi_a\ov x\,\xi_a +1}\,,
 \eal
and in the mirror kinematics we have
\bal
\tx_a(\tg)&={\xi_a-i\,e^{\tg} \ov 1 + i\,\xi_a e^{\tg} }\,,
\qquad
\tg_a(\tx) = \ln{\xi_a -\tx \ov \tx\,\xi_a +1}-\frac{i\pi}{2}\,.
 \eal
These expressions appear in the dressing factors through the function $R(\g)$,
\begin{equation}
    R (\g)= {G(1- \frac{\g}{2\pi i})\ov G(1+ \frac{\g}{2\pi i}) }\,,
\end{equation}
where $G(\g)$ is the Barnes $G$-function.

\paragraph{Functions $\alpha(x)$.} In the dressing factors we also use the function
\begin{equation}
    \tilde{\a}^{\pm m}_{a} = \alpha_a(\tx^{\pm m}_{a}(u))\,,\qquad\alpha_a(x)=\left(1-\frac{\xi_a}{x}\right)\left(x+\frac{1}{\xi_a}\right),
\end{equation}
not to be confused with the parameter $0<\alpha<1$.
Its properties (and those of $\sqrt{\tilde{\alpha}}$) under crossing are described in Appendix A of~\cite{Frolov:2025tda}.  For reader's convenience we repeat them below.

The function $\a_a(x)$ transforms as follows under the three discrete symmetries of the mirror region
\bal
 \alpha_a(1/x) =  -\alpha_{\bar a}(x)\,,\quad  \alpha_a(-x) =  -\alpha_{\bar a}(x)\,,\quad  \alpha_a(-1/x) =  \alpha_{a}(x)\,.\quad 
\eal
Then, taking into account that 
\bal
\sgn\left( \Im\left(\alpha_{a}(x)\right)\right) = \sgn\left( \Im(x)\right)\,,
\eal
we get that 
\bal
\sqrt{\a_a(-x)}= -\sgn\left( \Im(x)\right)\,i\,\sqrt{\a_{\bar a}(x)}\,,\quad \sqrt{\a_a(1/x)}= -\sgn\left( \Im(x)\right)\,i\,\sqrt{\a_{\bar a}(x)}\,,
\eal
Using that the cuts of $\sqrt{\a_a(x)}$ are the intervals $(-\infty, -1/\xi_a)$ and $(0, \xi_a)$,
we find that 
moving $x$ to $1/x$  through the cuts of $\sqrt{\a_a(x)}$ gives
\bal
\sqrt{\a_a(x)}\ \ \xrightarrow{x\to 1/x\ \text{through a cut}}\ \  +\sgn\left( \Im(x)\right)\,i\,\sqrt{\a_{\bar a}(x)}\,.
\eal

\paragraph{Dressing factors.}
The (modified) Beisert-Eden-Staudacher (BES) phase~\cite{Beisert:2006ez} is expressed in terms of a double integral~\cite{Dorey:2007xn}
\begin{equation}
    \tPhi_{ab}^{\alpha \beta}(x_{1},x_{2}) 
= -  \lint_{{ \pa\cR_\alpha}} \frac{{\rm d} w_1}{2\pi i} \lint_{{ \pa\cR_\beta}} \frac{{\rm d} w_2}{2\pi i} {1\ov w_1-x_{1}}{1\ov w_2-x_{2}} K^\bes_{(m)}(u_a(w_1)-u_b(w_2))\,.
\end{equation}
We refer the reader to~\cite{Frolov:2025uwz,Frolov:2025tda} for a discussion of the contours~$\pa\cR_\alpha$. Here we want to stress that we make used of the \textit{modified} BES Kernel
\begin{equation}
\label{eq:BESm}
    K^\bes_{(m)}(v)= i \log \frac{\Gamma \left(1+\frac{ih}{2m}v \right)}{\Gamma \left(1-\frac{ih}{2m}v \right)}\,,\qquad m=\alpha,\,1-\alpha\,,
\end{equation}
where $m=m_1=m_2$ is the mass of the particles being scattered. This definition cannot be straightforwardly generalised to the case $m_1\neq m_2$.
Expanding the BES Kernel at large-$h$ gives the Arutyunov-Frolov-Staudacher (AFS)~\cite{Arutyunov:2004vx} Kernel, the Hern\'andez-L\'opez (HL)~\cite{Hernandez:2006tk} Kernel, and subleading terms, like in~\cite{Frolov:2025uwz,Frolov:2025tda}, up to overall powers of~$m$. In particular, the HL phase is unchanged with repsect to that found in~\cite{Frolov:2025uwz,Frolov:2025tda}.

\section{Checking crossing and braiding unitarity}
\label{app:crossing}

To verify that the S-matrix elements satisfy the crossing equations and the braiding unitarity we use the following relations derived in \cite{Frolov:2025uwz}
\bal
 & \left(\frac{\Sigma^{\bes}_{aa}(\bar u_1, u_2)}{\Sigma^{\hl}_{ a a}(\bar u_1,u_2)}\right)^{-2}   \left(\frac{\Sigma^{\bes}_{\bar aa}( u_1, u_2)}{\Sigma^{\hl}_{\bar a a}( u_1,u_2)}\right)^{-2} =   \frac{u_1-u_2 - {2im\ov h}}{u_1-u_2 + {2im\ov h}}\,,
  \\
  &   \left(\frac{\Sigma^{\bes}_{\bar aa}( \bar u_1, u_2)}{\Sigma^{\hl}_{\bar a a}(\bar u_1,u_2)}\right)^{-2} \left(\frac{\Sigma^{\bes}_{aa}( u_1, u_2)}{\Sigma^{\hl}_{ a a}( u_1,u_2)}\right)^{-2} =   \frac{u_1-u_2 - {2im\ov h}}{u_1-u_2 + {2im\ov h}}\,,
\eal
where, as we mentioned before,  $h$ in the BES factors is rescaled as $h\to h/m$, and
\bal
\label{eq:RidentityGen}
 & R (\g \pm2\pi i) = i\left({ \sinh{\g\ov2}\ov \pi }\right)^{\pm1}R(\g)\,,\qquad R (\g+\pi i) =  { \cosh{\g\ov2}\ov \pi }R(\g-\pi i)\,,
\eal
with $R (-\g) R(\g)=1$. From this and from the definitions in appendix~\ref{app:def} one can derive the following useful relations:
\bal
\frac{R(\tg^{-+}_{ \R\L}- i \pi) R(\tg^{+-}_{ \R\L}+ i \pi)}{R(\tg^{--}_{ \R\L}- i \pi) R(\tg^{++}_{\R\L}+ i \pi)} &= {\cosh{\tg^{+-}_{ \R\L}\ov2} \ov \cosh{\tg^{++}_{ \R\L}\ov2}  }\frac{R(\tg^{-+}_{ \R\L}- i \pi) R(\tg^{+-}_{ \R\L}- i \pi)}{R(\tg^{--}_{ \R\L}- i \pi) R(\tg^{++}_{\R\L}- i \pi)}  
\\
&={\sqrt{\tilde\a_{\L2}^{+}}\ov \sqrt{\tilde\a_{\L2}^{-}}}\,{\sqrt{\tx_{\L2}^{+}}\ov \sqrt{\tx_{\L2}^{-}}}\,{ 1-\tx^{+}_{\R1}\tx^{-}_{\L2}\ov1-\tx^{+}_{\R1}\tx^{+}_{\L2}} \frac{R(\tg^{-+}_{ \R\L}- i \pi) R(\tg^{+-}_{ \R\L}- i \pi)}{R(\tg^{--}_{ \R\L}- i \pi) R(\tg^{++}_{\R\L}- i \pi)}  
\eal
\bal\label{eq:Ridentity}
\frac{R(\tg^{-+}_{ \R\L}- i \pi) R(\tg^{+-}_{ \R\L}+ i \pi)}{R(\tg^{--}_{ \R\L}- i \pi) R(\tg^{++}_{\R\L}+ i \pi)} &= {\cosh{\tg^{--}_{ \R\L}\ov2} \ov \cosh{\tg^{-+}_{ \R\L}\ov2}  }\frac{R(\tg^{-+}_{ \R\L}+ i \pi) R(\tg^{+-}_{ \R\L}+ i \pi)}{R(\tg^{--}_{ \R\L}+ i \pi) R(\tg^{++}_{\R\L}+ i \pi)}  
\\
&={\sqrt{\tilde\a_{\L2}^{+}}\ov \sqrt{\tilde\a_{\L2}^{-}}}\,{\sqrt{\tx_{\L2}^{+}}\ov \sqrt{\tx_{\L2}^{-}}}\,{ 1-\tx^{-}_{\R1}\tx^{-}_{\L2}\ov1-\tx^{-}_{\R1}\tx^{+}_{\L2}} \frac{R(\tg^{-+}_{ \R\L}+i \pi) R(\tg^{+-}_{ \R\L}+ i \pi)}{R(\tg^{--}_{ \R\L}+ i \pi) R(\tg^{++}_{\R\L}+ i \pi)}  
\eal

\bal
\frac{R(\tg^{-+}_{ \R\L}- i \pi) R(\tg^{+-}_{ \R\L}+ i \pi)}{R(\tg^{--}_{ \R\L}- i \pi) R(\tg^{++}_{\R\L}+ i \pi)} &={{\tilde\a_{\L2}^{+}}\ov {\tilde\a_{\L2}^{-}}}\,{{\tx_{\L2}^{+}}\ov {\tx_{\L2}^{-}}}\,{ 1-\tx^{-}_{\R1}\tx^{-}_{\L2}\ov1-\tx^{-}_{\R1}\tx^{+}_{\L2}}\,{ 1-\tx^{+}_{\R1}\tx^{-}_{\L2}\ov1-\tx^{+}_{\R1}\tx^{+}_{\L2}} 
 \frac{R(\tg^{-+}_{ \R\L}+i \pi) R(\tg^{+-}_{ \R\L}- i \pi)}{R(\tg^{--}_{ \R\L}+ i \pi) R(\tg^{++}_{\R\L}- i \pi)}  
\eal

Making use of the above relations, it is easy to check the crossing equations
\bal
   S^{m m}_{\cX\cX}(\bar u_1,u_2)&S^{mm}_{\bar{\Upsilon}\mathcal{X}} (u_1,u_2)=\\
   &=H^{mm}_{\cX\cX}  (\bar u_1,u_2)H^{mm}_{\bar{\Upsilon}\mathcal{X}}
(u_1,u_2)
{\sqrt{\tilde\a_{\L2}^{+m}}\ov \sqrt{\tilde\a_{\L2}^{-m}}}
    \frac{ R(\tg^{-m-m}_{\R\L}-i\pi)}{R(\tg^{-m+m}_{\R\L} -i\pi)} \frac{R(\tg^{-m+m}_{ \R\L}+ i \pi) }{R(\tg^{-m-m}_{ \R\L}+ i \pi)}
    \\
    &=H^{mm}_{\cX\cX}  (\bar u_1,u_2)H^{mm}_{\bar{\Upsilon}\mathcal{X}}(u_1,u_2)\,g(\tx_{\R1}^{\pm m},\tx_{\L2}^{\pm m})\,,
   \eal
and
\bal
S^{mm}_{\bar{\Upsilon}\mathcal{X}} (\bar u_1,u_2)   S^{m m}_{\cX\cX}( u_1,u_2)& =H^{mm}_{\bar{\Upsilon}\mathcal{X}}(\bar u_1,u_2)   H^{mm}_{\cX\cX}  ( u_1,u_2){\sqrt{\tilde\a_{\L2}^{+m}}\ov \sqrt{\tilde\a_{\L2}^{-m}}} \frac{R(\tg^{+m-m}_{ \L\L}- 2i \pi)}{R(\tg^{+m+m}_{\L\L}-2 i \pi)} \frac{R(\tg^{+m+m}_{\L\L}) }{R(\tg^{+m-m}_{\L\L})  }
  \\
    &=H^{mm}_{\bar{\Upsilon}\mathcal{X}}(\bar u_1,u_2)   H^{mm}_{\cX\cX}  ( u_1,u_2)\,f(\tx_{\L1}^{\pm m},\tx_{\L2}^{\pm m})\,.
   \eal
With similar manipulations we find
   \bal 
      &S^{m m}_{\Upsilon\Upsilon}(\bar u_1,u_2)S^{mm}_{\bar\cX\Upsilon}(u_1,u_2)=
    -H^{mm}_{\cX\cX} (\bar u_1,u_2)H^{mm}_{\bar{\Upsilon}\mathcal{X}}(u_1,u_2)\,\tilde g(\tx_{\R1}^{\pm m},\tx_{\L2}^{\pm m})\,,\\
      &S^{mm}_{\bar\cX\Upsilon}(\bar u_1,u_2)S^{m m}_{\Upsilon\Upsilon}( u_1,u_2)=
    -H^{mm}_{\bar{\Upsilon}\mathcal{X}}(\bar u_1,u_2)H^{mm}_{\cX\cX} ( u_1,u_2)\,\tilde f(\tx_{\R1}^{\pm m},\tx_{\L2}^{\pm m})\,,
    \eal
    and, therefore,  we should require that $ H^{mm}_{\cX\cX} $ and $H^{mm}_{\bar{\Upsilon}\mathcal{X}}$ satisfy the homogeneous crossing equations
   \bal
  H^{mm}_{\cX\cX}  (\bar u_1,u_2)H^{mm}_{\bar{\Upsilon}\mathcal{X}}(u_1,u_2)=1\,,\qquad   H^{mm}_{\bar{\Upsilon}\mathcal{X}}(\bar u_1,u_2)   H^{mm}_{\cX\cX}  ( u_1,u_2)=1\,.
\eal 
The crossing equations for different-mass crossing are checked by using exactly the same formulae.

The braiding unitarity for $S^{m_1 m_2}_{\cX\cX}(\bar u_1,u_2)$ and $S^{m_1 m_2}_{\Upsilon\Upsilon}(\bar u_1,u_2)$ for any choice of $m_1,m_2$ is obvious. For the other S-matrix elements we get, using the $R$-function relations above,
\bal
&S^{mm}_{\bar{\Upsilon}\mathcal{X}} ( u_1,u_2)   S^{m m}_{\mathcal{X}\bar{\Upsilon}}( u_2,u_1)=\\
&=H^{mm}_{\bar{\Upsilon}\mathcal{X}}( u_1,u_2)   H^{mm}_{{\Upsilon}\bar\cX}  ( u_2,u_1){\sqrt{\tilde\a_{\R1}^{-m}}\ov \sqrt{\tilde\a_{\R1}^{+m}}} {\sqrt{\tilde\a_{\L2}^{+m}}\ov \sqrt{\tilde\a_{\L2}^{-m}}}{\sqrt{\tx_{\R1}^{-m}}\ov \sqrt{\tx_{\R1}^{+m}}}\, {\sqrt{\tx_{\L2}^{+m}}\ov \sqrt{\tx_{\L2}^{-m}}}{1-{\tx^{+m}_{\R1}} \tx^{-m}_{\L2}\ov  1-{ \tx^{-m}_{\R1}}\tx^{+m}_{\L2}}
\\
&\qquad\times  {R(\tg^{-m+m}_{ \R\L}+ i \pi) R(\tg^{+m-m}_{ \R\L}- i \pi)}
{R(-\tg^{+m-m}_{ \R\L}- i \pi) R(-\tg^{-m+m}_{ \R\L}+ i \pi)}
\\
  \\
&=H^{mm}_{\bar{\Upsilon}\mathcal{X}}( u_1,u_2)   H^{mm}_{{\Upsilon}\bar\cX}  ( u_2,u_1) =1
\,,
   \eal
   and, similarly, 
\bal
S^{m_1m_2}_{\bar{\Upsilon}\mathcal{X}} ( u_1,u_2)  & S^{m_2 m_1}_{\mathcal{X}\bar{\Upsilon}}( u_2,u_1)=H^{m_1m_2}_{\bar{\Upsilon}\mathcal{X}}( u_1,u_2)   H^{m_2m_1}_{{\Upsilon}\bar\cX}  ( u_2,u_1) =1\,.
    \eal   
    Clearly, the function \eqref{eq:Hm1m2} satisfy the homogeneous crossing equations and the braiding unitarity.

\section{The BMN limit}
\label{app:BMN}
Here we expand our proposal in the near BMN limit and find the following expansion
\bal
\mathcal{S}=1+ i \, \mathcal{T}_{0}\,,
\eal
where the $\mathcal{T}_0$ operator is computed in the uniform lightcone gauge with parameter $a=0$~\cite{Arutyunov:2009ga}. To compare with perturbative computations such as the ones in~\cite{Bianchi:2014rfa} we need this function  in the gauge $a=1/2$, given by
\bal
\mathcal{T}_{1/2}=\mathcal{T}_{0} + \frac{\tilde{\omega}_1 \tp_2 - \tilde{\omega}_2 \tp_1}{2} \,.
\eal

\subsection{Equal masses}

The limit for equal masses is similar to the one performed in appendix J.1 of~\cite{Frolov:2025uwz}. It is sufficient to consider one S-matrix element for the right-right scattering and one S-matrix element for the right-left scattering because the others are fixed by symmetries. 

\paragraph{Right-Right Scattering.} It is convenient to consider 
$e^{{i\ov2}({\tilde{\omega}_{\R1} \tp_2 - \tilde{\omega}_{\R2} \tp_1})}S^{mm}_{\bar\psi\bar\psi}$. It corresponds to the S-matrix element $-F_{++}(p,p')$ in ~\cite{Bianchi:2014rfa}, and in the pure RR case it is equal to $-1$ at the AFS order, see eqs. (3.10), (3.12) in~\cite{Bianchi:2014rfa}.

It is given by 
\bal\label{eq:Sbpsibpsimm}
 - e^{{i\ov2}({\tilde{\omega}_{\R1} \tp_2 - \tilde{\omega}_{\R2} \tp_1})} S^{m m}_{\bar\psi\bar\psi}(u_1,u_2)=&\cA(u_1,u_2)\,\cB(u_1,u_2)\,\cR(u_1,u_2)\,,
\eal
where
\bal
   \cA(u_1,u_2)=e^{{i\ov2}({\tilde{\omega}_{\R1} \tp_2 - \tilde{\omega}_{\R2} \tp_1})}H^{mm}_{\bar\cX\bar\cX}  (u_1,u_2)\,  {\sqrt{\tx_{\L1}^{+m}}\ov \sqrt{\tx_{\L1}^{-m}}}\, {\sqrt{\tx_{\L2}^{-m}}\ov \sqrt{\tx_{\L2}^{+m}}}\, { \tx^{-m}_{\L1}- \tx^{+m}_{\L2}\ov \tx^{+m}_{\L1}- \tx^{-m}_{\L2}}
   \,,
    \eal
\bal
   \cB(u_1,u_2)=
    \frac{u_1-u_2 + {2im\ov h}}{u_1-u_2 - {2im\ov h}}\,
    \left(\frac{\Sigma^{\bes}_{\L\L}(\tx^{\pm m}_{\L1}, \tx^{\pm m}_{\L2})}{\Sigma^{\hl}_{\L\L}(\tx^{\pm m}_{\L1}, \tx^{\pm m}_{\L2})}\right)^{-2}\,,
    \eal    
\bal
   \cR(u_1,u_2)=
    \frac{R(\tg^{+m+m}_{\L\L}) R(\tg^{-m-m}_{\L\L})}{R(\tg^{+m-m}_{\L\L}) R(\tg^{-m+m}_{\L\L}) }\,.
    \eal
The odd part%
\footnote{Following~\cite{Beisert:2006ib} we call ``odd'' the part of the dressing factor which gives a net contribution to the double-crossing equation.}
$\cR(u_1,u_2)$ given by the ratios of the R-functions does not contribute to the AFS order. 
We expand the Zhukovsky variables as follows
\bal
\tx^{\pm m}_{\R i} &= \tx_{\R i} \pm \frac{i}{h'} \frac{1}{u'_{\R}(\tx_{\R i})} 
\,,\qquad h' = \frac{h}{m}\,.
\eal
and use that in the large $h$ limit 
\bal
\label{eq:RR_mir_mir_BMN_1}
\frac{1}{i} \log \cB(u_1,u_2) &\to -\frac{1}{h'} f_x(\tx_{\R1}, \tx_{\R2})
\eal
where the function $f_x$ is given in equation (J.9) of~\cite{Frolov:2025uwz}
\bal
\label{eq:fx_defined}
f_{x}(\tx_{\R1}, \tx_{\R2})\equiv& - 2\frac{u'_\R(\tx_{\R1})+u'_\R(\tx_{\R2})}{u'_\R(\tx_{\R1}) u'_\R(\tx_{\R2})} \frac{1}{\tx_{\R1} - \tx_{\R2}} + \frac{1}{\tx_{\R1} u'_\R(\tx_{\R1})} - \frac{1}{\tx_{\R2} u'_\R(\tx_{\R2})}\\
&- \frac{1}{u'_\R(\tx_{\R1}) u'_\R(\tx_{\R2})}\left( \frac{1}{\tx_{\R1}} -  \frac{1}{\tx_{\R2}}+ \frac{1}{\tx_{\R1} \tx^2_{\R2}} -  \frac{1}{\tx^2_{\R1} \tx_{\R2}} \right)\,.
\eal
Taking into account that the momentum in terms of the Zhukovsky variables is given by
\bal
\tp_{\R i} = \frac{h}{2} \left( \tx^{-m}_{\R i} - \frac{1}{\tx^{-m}_{\R i}} -\tx^{+m}_{\R i} -\frac{1}{\tx^{+m}_{\R i}} \right) = - \frac{i m \sqrt{1-q^2} (\tx^2_{\R i} +1)}{\sqrt{1-q^2} (\tx^2_{\R i} -1) + 2q \tx_{\R i}} + \mathcal{O}(T^{-2}) \,,
\eal
one gets
\bal
\tx_{\R i} = \frac{\sqrt{\tp^2_{i} + m^2 (1-q^2) } - q \tp_{i}}{(\tp_{i} + i m) \sqrt{1-q^2}} \,.
\eal
Substituting this expression into the function $f_x(\tx_{\R1}, \tx_{\R2})$, we find
\bal
\label{eq:f_evaluated}
f_x(\tx_{\R1}, \tx_{\R2})= - \frac{\sqrt{1-q^2}}{m^2} \, \tilde{\omega}_{\R1} \, \tilde{\omega}_{\R2} \, \frac{\tp_{1} + \tp_{2}}{\tilde{\omega}_{\R1} - \tilde{\omega}_{\R2}}\,,
\eal
where we defined 
\bal
\tilde{\omega}_{\R i} = - i q m + \sqrt{\tp^2_{i} + m^2 (1- q^2)}\,.
\eal
Next, substituting \eqref{eq:f_evaluated} into~\eqref{eq:RR_mir_mir_BMN_1} we obtain
\bal
\frac{1}{i} \log \cB(u_1,u_2)  &\to \frac{1}{m T} \tilde{\omega}_{\R 1} \tilde{\omega}_{\R 2} \frac{\tp_{1} + \tp_{2}}{\tilde{\omega}_{\R 1}-\tilde{\omega}_{\R 2}} \,.
\eal
The large $h$ expansion of $\cA$ is straightforward, and, summing up the two contributions, we get 0.
This matches  with the perturbative results from~\cite{Bianchi:2014rfa}, and, moreover, $\mathcal{T}_{\bar{\psi} \bar{\psi}}$ vanishes for any $q$ at the AFS order.

\paragraph{Right-Left Scattering.} Here we check that the near BMN expansion of the right-left scattering elements for particles of equal mass agree with the known perturbative results. In this case it is convenient to consider 
$e^{{i\ov2}({\tilde{\omega}_{\R1} \tp_2 - \tilde{\omega}_{\L2} \tp_1})}S^{mm}_{\bar\psi\psi}$. It corresponds to the S-matrix element $E_{+-}(p,p')$ in ~\cite{Bianchi:2014rfa}, and in the pure RR case it is equal to $+1$ at the AFS order, see eqs. (3.11), (3.13) in~\cite{Bianchi:2014rfa}.

The S-matrix element $S^{mm}_{\bar\psi\psi}$ is related to $S^{mm}_{\bar X\psi}$ in \eqref{eq:SYYmm} as follows
\bal
S^{mm}_{\bar\psi\psi}= {\sqrt{\tx_{\L2}^{-m}}\ov \sqrt{\tx_{\L2}^{+m}}} \,{{1- \tx^{-m}_{\R1}\tx^{+m}_{\L2}}\ov  {1-\tx^{-m}_{\R1}\tx^{-m}_{\L2}}}\, S^{mm}_{\bar X\psi}(u_1,u_2)
 \,.
\eal
We  use identity \eqref{eq:Ridentity} to bring $S^{mm}_{\bar X\psi}$ to the form
\bal
S^{mm}_{\bar X\psi}(u_1,u_2)=&+H^{mm}_{\bar{\Upsilon}\mathcal{X}}(u_1,u_2)\,
{\sqrt{\tx_{\R1}^{-m}}\ov \sqrt{\tx_{\R1}^{+m}}}\, {{\tx_{\L2}^{+m}}\ov {\tx_{\L2}^{-m}}} \,{1- \tx^{+m}_{\R1} \tx^{-m}_{\L2}\ov  1-\tx^{-m}_{\R1} \tx^{+m}_{\L2}}\, {1- \tx^{-m}_{\R1} \tx^{-m}_{\L2}\ov  1-\tx^{-m}_{\R1} \tx^{+m}_{\L2}}\, 
\\
&\quad\times \frac{R(\tg^{-+}_{ \R\L}+i \pi) R(\tg^{+-}_{ \R\L}+ i \pi)}{R(\tg^{--}_{ \R\L}+ i \pi) R(\tg^{++}_{\R\L}+ i \pi)}    \left(\frac{\Sigma^{\bes}_{\R\L}(\tx^{\pm m}_{\R1}, \tx^{\pm m}_{\L2})}{\Sigma^{\hl}_{\R\L}(\tx^{\pm m}_{\R1}, \tx^{\pm m}_{\L2})}\right)^{-2} \,.
\eal
Then, we get
\bal
S^{mm}_{\bar\psi\psi}&=H^{mm}_{\bar{\Upsilon}\mathcal{X}}(u_1,u_2)\,
{\sqrt{\tx_{\R1}^{-m}}\ov \sqrt{\tx_{\R1}^{+m}}}\,  {\sqrt{\tx_{\L2}^{+m}}\ov \sqrt{\tx_{\L2}^{-m}}} \,{1- \tx^{+m}_{\R1} \tx^{-m}_{\L2}\ov  1-\tx^{-m}_{\R1} \tx^{+m}_{\L2}}
\\
&\quad\times \frac{R(\tg^{-+}_{ \R\L}+i \pi) R(\tg^{+-}_{ \R\L}+ i \pi)}{R(\tg^{--}_{ \R\L}+ i \pi) R(\tg^{++}_{\R\L}+ i \pi)}    \left(\frac{\Sigma^{\bes}_{\R\L}(\tx^{\pm m}_{\R1}, \tx^{\pm m}_{\L2})}{\Sigma^{\hl}_{\R\L}(\tx^{\pm m}_{\R1}, \tx^{\pm m}_{\L2})}\right)^{-2} 
 \,.
\eal
The advantage of this form is that the R-function part does not contribute  to the AFS order. The large $h$ expansion of the left Zhukovsky variables is given by the same formulae as for the right ones with the replacement $q\to -q$.

The expansion of the BES can be found by using (C.11) in~\cite{Frolov:2025uwz}, and is given by
\bal
{1\ov i}\log\left(\frac{\Sigma^{\bes}_{\R\L}(\tx^{\pm m}_{\R1}, \tx^{\pm m}_{\L2})}{\Sigma^{\hl}_{\R\L}(\tx^{\pm m}_{\R1}, \tx^{\pm m}_{\L2})}\right)^{-2} = -8 \Delta \tx_{\R1}\Delta \tx_{\L2} {\stackrel{\prime\prime}{\tPhi}}{}^{\afs}_{\R\L}(\tx_{\R1},\tx_{\L2}) 
\eal
where
\bal
{\stackrel{\prime\prime}{\tPhi}}{}^{\afs}_{\R\L}(\tx_{\R1},\tx_{\L2}) = h'\,\frac{\tx_{\R1}-\tx_{\L2}+{k\ov 2 \pi h} (\tx_{\R1} \tx_{\L2}+1)}{2 \tx_{\R1} \tx_{\L2} (\tx_{\R1} \tx_{\L2}-1)}\,,
\eal
and 
\bal
\Delta\tx_{\R1} &=\frac{i}{h'} \frac{1}{u'_{\R}(\tx_{\R1})} 
\,,\qquad \Delta\tx_{\L2} &=\frac{i}{h'} \frac{1}{u'_{\L}(\tx_{\L2})}\,.
\eal
Performing the large $h$ expansion, we indeed find that for any $q$
\bal
{1\ov i}\log\left( e^{{i\ov2}({\tilde{\omega}_{\R1} \tp_2 - \tilde{\omega}_{\L2} \tp_1})}S^{mm}_{\bar\psi\psi} \right) = \cO(T^{-2})\,,
\eal
in agreement with \cite{Bianchi:2014rfa}.

\subsection{Different masses}

The S-matrix elements involving one particle of mass $\a$ and one of mass $1-\a$ do not involve the BES factor (which we would  not know how to define in this case). For the right-right scattering the convenient S-matrix element to consider is 
$e^{{i\ov2}({\tilde{\omega}_{\R1} \tp_2 - \tilde{\omega}_{\R2} \tp_1})}S^{m_1m_2}_{\bar X\bar X}$. It corresponds to the S-matrix element $A_{++}(p,p')$ in ~\cite{Bianchi:2014rfa}, and in the pure RR case it is equal to $1$ at the AFS order, see eqs. (3.15), (3.18) in~\cite{Bianchi:2014rfa}. Since $e^{{i\ov2}({\tilde{\omega}_{\R1} \tp_2 - \tilde{\omega}_{\R2} \tp_1})}S^{m_1m_2}_{\bar X\bar X}$ is just given by the product of R-functions, its large $h$ expansion starts at order $1/T^2$ in agreement with \cite{Bianchi:2014rfa}.

Similarly, for the right-left scattering we consider the element 
$e^{{i\ov2}({\tilde{\omega}_{\R1} \tp_2 - \tilde{\omega}_{\L2} \tp_1})}S^{m_1m_2}_{\bar X  X}$. It corresponds to the S-matrix element $A_{+-}(p,p')$ in ~\cite{Bianchi:2014rfa}, and in the pure RR case it is also equal to $1$ at the AFS order, see eqs. (3.16), (3.19) in~\cite{Bianchi:2014rfa}. Since $e^{{i\ov2}({\tilde{\omega}_{\R1} \tp_2 - \tilde{\omega}_{\L2} \tp_1})}S^{m_1m_2}_{\bar X X}$ is again given by a product of R-functions, its large $h$ expansion starts at order $1/T^2$ as well in agreement with \cite{Bianchi:2014rfa}.

\section{Other solutions when \texorpdfstring{$\alpha=1/2$}{alpha=1/2}}
\label{app:solutions}
One of the difficulties in proposing the dressing factors for general values of~$\alpha$ is that it is unclear whether the BES factor admits a generalisation when scattering excitations of different masses (which are not multiples of each other). This is not an issue when $\alpha=1/2$, where it is easy to construct more solutions.  Both solutions in this appendix have the right pole structure and agree with available perturbative results.

\subsection{BES dressing factor in all processes}
\label{app:solutions:1}
In this case $m_1= m_2 = 1/2$, and we keep \eqref{eq:SXXmm} and  \eqref{eq:SYYmm} unchanged but we modify \eqref{eq:SXXm1m2} and  \eqref{eq:SYYm1m2} as follows
\bal
\label{eq:SXXm1m2b}
S^{m_1m_2}_{\cX\cX} (u_1,u_2)&=H^{mm}_{\cX\cX} (u_1,u_2)\, {\sqrt{\tx_{\L1}^{+m}}\ov \sqrt{\tx_{\L1}^{-m}}}\, {\sqrt{\tx_{\L2}^{-m}}\ov \sqrt{\tx_{\L2}^{+m}}}\, { \tx^{-m}_{\L1}- \tx^{+m}_{\L2}\ov \tx^{+m}_{\L1}- \tx^{-m}_{\L2}}
    \frac{u_{12} + {2im\ov h}}{u_{12} - {2im\ov h}}\,
    \\
    &\quad\times
    \frac{R(\tg^{+m+m}_{\L\L}) R(\tg^{-m-m}_{\L\L})}{R(\tg^{+m-m}_{\L\L}) R(\tg^{-m+m}_{\L\L}) }
     \left(\frac{\Sigma^{\bes}_{\L\L}(\tx^{\pm m}_{\L1}, \tx^{\pm m}_{\L2})}{\Sigma^{\hl}_{\L\L}(\tx^{\pm m}_{\L1}, \tx^{\pm m}_{\L2})}\right)^{-2}\,,
\\
 S^{m_1m_2}_{\bar{\Upsilon}\mathcal{X}} (u_1,u_2)&=H^{mm}_{\bar{\Upsilon}\mathcal{X}}
(u_1,u_2)\, {\sqrt{\tilde\a_{\L2}^{+m}}\ov \sqrt{\tilde\a_{\L2}^{-m}}}\,{\sqrt{\tx_{\R1}^{+m}}\ov \sqrt{\tx_{\R1}^{-m}}}\, {\sqrt{\tx_{\L2}^{+m}}\ov \sqrt{\tx_{\L2}^{-m}}}  \,{{1\ov \tx^{+m}_{\R1}}- \tx^{-m}_{\L2}\ov  {1\ov\tx^{-m}_{\R1}} - \tx^{+m}_{\L2}}\,
\\&
\times\frac{R(\tg^{-m+m}_{ \R\L}+ i \pi) R(\tg^{+m-m}_{ \R\L}- i \pi)}
{R(\tg^{-m-m}_{ \R\L}+ i \pi) R(\tg^{+m+m}_{\R\L}- i \pi)}  \left(\frac{\Sigma^{\bes}_{\R\L}(\tx^{\pm m}_{\R1}, \tx^{\pm m}_{\L2})}{\Sigma^{\hl}_{\R\L}(\tx^{\pm m}_{\R1}, \tx^{\pm m}_{\L2})}\right)^{-2}\,,
\eal
where $u_{12}=u_1-u_2$, and 
\bal
\label{eq:SYYm1m2b}
S^{m_1m_2}_{{\Upsilon}{\Upsilon}} (u_1,u_2)=&-H^{mm}_{\cX\cX} (u_1,u_2)\ \frac{u_{12} + {2im\ov h}}{u_{12} - {2im\ov h}}\, 
\frac{R(\tg^{+m+m}_{\L\L}) R(\tg^{-m-m}_{\L\L})}{R(\tg^{+m-m}_{\L\L}) R(\tg^{-m+m}_{\L\L}) } \left(\frac{\Sigma^{\bes}_{\L\L}(\tx^{\pm m}_{\L1}, \tx^{\pm m}_{\L2})}{\Sigma^{\hl}_{\L\L}(\tx^{\pm m}_{\L1}, \tx^{\pm m}_{\L2})}\right)^{-2},\\
S^{m_1m_2}_{\bar\cX\Upsilon}(u_1,u_2)=&H^{mm}_{\bar{\Upsilon}\mathcal{X}}
(u_1,u_2)\,\frac{R(\tg^{-m+m}_{ \R\L}- i \pi) R(\tg^{+m-m}_{ \R\L}+ i \pi)}{R(\tg^{-m-m}_{ \R\L}- i \pi) R(\tg^{+m+m}_{\R\L}+ i \pi)} \left(\frac{\Sigma^{\bes}_{\R\L}(\tx^{\pm m}_{\R1}, \tx^{\pm m}_{\L2})}{\Sigma^{\hl}_{\R\L}(\tx^{\pm m}_{\R1}, \tx^{\pm m}_{\L2})}\right)^{-2} .
\eal
Note that on the left-hand-side we kept the notation $S^{m_1m_2}$ to highlight that these processes are related to the scattering of excitations arising from \textit{different spheres}. However, since $m_1=m_2=m=1/2$, we have simplified the right-hand-side by using $m$ only.
Recall that
\bal
      S^{m m}_{\Upsilon\Upsilon}(u_1,u_2)=&-H^{mm}_{\cX\cX} (u_1,u_2)\, {\sqrt{\tx_{\L1}^{+m}}\ov \sqrt{\tx_{\L1}^{-m}}}\, {\sqrt{\tx_{\L2}^{-m}}\ov \sqrt{\tx_{\L2}^{+m}}}\, { \tx^{-m}_{\L1}- \tx^{+m}_{\L2}\ov \tx^{+m}_{\L1}- \tx^{-m}_{\L2}}
    \frac{u_{12} + {2im\ov h}}{u_{12} - {2im\ov h}}\,
    \\
    &\quad\times
    \frac{R(\tg^{+m+m}_{\L\L}) R(\tg^{-m-m}_{\L\L})}{R(\tg^{+m-m}_{\L\L}) R(\tg^{-m+m}_{\L\L}) }
     \left(\frac{\Sigma^{\bes}_{\L\L}(\tx^{\pm m}_{\L1}, \tx^{\pm m}_{\L2})}{\Sigma^{\hl}_{\L\L}(\tx^{\pm m}_{\L1}, \tx^{\pm m}_{\L2})}\right)^{-2}
   \,,
\eal
and therefore (note that the physical S-matrix differs from $ S^{m m}_{\Upsilon\Upsilon}$ by a sign, in the sense of~\cite{Frolov:2025ozz})
\bal
S^{m_1m_2}_{\cX\cX} (u_1,u_2) = -   S^{m m}_{\Upsilon\Upsilon}(u_1,u_2)\,.
\eal

\paragraph{Issues with fusion.} 
Aside from the fact that it is not obvious how to define the BES factor for different masses when $\alpha\neq1/2$ (and hence how to generalise this proposal), we note that this proposal presents an ambiguity when performing fusion. Consider the ``strange Bethe string'' of~\eqref{eq:rightfusion}. This is still a valid configuration in the thermodynamic limit, because we have not changed the pole structure of the model. However, we now have the fused S-matrix element
\begin{equation}
\begin{aligned}
& S^{m_1 m_1}_{\Upsilon\Upsilon}(u_1,v)\, S^{m_1 m_1}_{\Upsilon\Upsilon}(u_1',v)\, S^{m_2 m_1}_{\Upsilon\Upsilon}(u_2,v)\, S^{m_2 m_1}_{\Upsilon\Upsilon}(u_2',v)
\\
&= {\sqrt{\tx_{\L}(u+{i\ov h})}\ov \sqrt{\tx_{\L}(u-{i\ov h})}}\, {{\tx_{\L}^{-m}(v)}\ov {\tx_{\L}^{+m}(v)}}\, { \tx_{\L}(u)- \tx_{\L}^{+m}(v)\ov \tx_{\L}(u+{i\ov h}) - \tx^{-m}_{\L}(v)} \, { \tx_{\L}(u-{i\ov h})- \tx_{\L}^{+m}(v)\ov \tx_{\L}(u)- \tx^{-m}_{\L}(v)} \times \ldots\,,
\end{aligned}
\end{equation}
where we used that $m_1=m_2=m=1/2$. This expression is well-defined when~$u$ is complex, but physical unitarity requires $u$ to be real. However, $\tx_{\L}(u)$ may have a cut there. To preserve unitarity we would have to require that 
\bal
u_1 = u +{i\ov 2h} - i0\,,\quad u_1' = u +{i\ov 2h} + i0\qquad \text{or}\quad  u_1 = u +{i\ov 2h} + i0\,,\quad u_1' = u +{i\ov 2h} - i0\,,
\eal
but even then the result would depend on a choice of the sign of $i0$. It is worth remarking that there is no such a problem with our main proposal of section~\ref{sec:proposal}.

\subsection{BES dressing factor for different-mass scattering only}
\label{app:solutions:2}

One can find a solution of the crossing equations which has BES/HL in the S-matrix elements  of particles from different spheres. Taking into account that $S^{m m}_{\Upsilon\Upsilon}$ is equal to 0 at the tree-level in the $a=1/2$ gauge, we get 
\bal\label{eq:SYYhh}
      S^{m m}_{\Upsilon\Upsilon}(u_1,u_2)=&-H^{mm}_{\cX\cX} (u_1,u_2)\, 
    \frac{R(\tg^{+m+m}_{\L\L}) R(\tg^{-m-m}_{\L\L})}{R(\tg^{+m-m}_{\L\L}) R(\tg^{-m+m}_{\L\L}) }
    \,,
    \\
S^{mm}_{\bar\cX\Upsilon}(u_1,u_2)=&+H^{mm}_{\bar{\Upsilon}\mathcal{X}}(u_1,u_2)\,{\sqrt{\tilde\a_{\L2}^{-m}}\ov \sqrt{\tilde\a_{\L2}^{+m}}}\,
\frac{R(\tg^{-m+m}_{ \R\L}- i \pi) R(\tg^{+m-m}_{ \R\L}+ i \pi)}{R(\tg^{-m-m}_{ \R\L}- i \pi) R(\tg^{+m+m}_{\R\L}+ i \pi)} \,,
\eal
\bal\label{eq:SXXhh}
   S^{m m}_{\cX\cX}(u_1,u_2)=&+H^{mm}_{\cX\cX}  (u_1,u_2)\,  {\sqrt{\tx_{\L1}^{+m}}\ov \sqrt{\tx_{\L1}^{-m}}}\, {\sqrt{\tx_{\L2}^{-m}}\ov \sqrt{\tx_{\L2}^{+m}}}\, { \tx^{-m}_{\L1}- \tx^{+m}_{\L2}\ov \tx^{+m}_{\L1}- \tx^{-m}_{\L2}}
    \frac{R(\tg^{+m+m}_{\L\L}) R(\tg^{-m-m}_{\L\L})}{R(\tg^{+m-m}_{\L\L}) R(\tg^{-m+m}_{\L\L}) }
   \,,
  \\
 S^{mm}_{\bar{\Upsilon}\mathcal{X}} (u_1,u_2)=&+H^{mm}_{\bar{\Upsilon}\mathcal{X}}
(u_1,u_2)\, {\sqrt{\tilde\a_{\L2}^{-m}}\ov \sqrt{\tilde\a_{\L2}^{+m}}}\, {\sqrt{\tx_{\R1}^{-m}}\ov \sqrt{\tx_{\R1}^{+m}}}\, {\sqrt{\tx_{\L2}^{-m}}\ov \sqrt{\tx_{\L2}^{+m}}}\,{1-{\tx^{+m}_{\R1}} \tx^{+m}_{\L2}\ov  1-{ \tx^{-m}_{\R1}}\tx^{-m}_{\L2}}\
\\
&\quad\times \frac{R(\tg^{-m+m}_{ \R\L}- i \pi) R(\tg^{+m-m}_{ \R\L}+ i \pi)}{R(\tg^{-m-m}_{ \R\L}- i \pi) R(\tg^{+m+m}_{\R\L}+ i \pi)} \,,
\eal
Note that  
\bal\label{eq:SYYhh2}
S^{m m}_{\cX\cX}(u_1,u_2)&=   S^{m m}_{\Upsilon\Upsilon}(u_1,u_2)\,   {\sqrt{\tx_{\L1}^{+m}}\ov \sqrt{\tx_{\L1}^{-m}}}\, {\sqrt{\tx_{\L2}^{-m}}\ov \sqrt{\tx_{\L2}^{+m}}}\, { \tx^{-m}_{\L1}- \tx^{+m}_{\L2}\ov \tx^{+m}_{\L1}- \tx^{-m}_{\L2}}
    \,,
    \\
 S^{mm}_{\bar{\Upsilon}\mathcal{X}} (u_1,u_2)&=S^{mm}_{\bar\cX\Upsilon}(u_1,u_2)\, {\sqrt{\tx_{\R1}^{-m}}\ov \sqrt{\tx_{\R1}^{+m}}}\, {\sqrt{\tx_{\L2}^{-m}}\ov \sqrt{\tx_{\L2}^{+m}}}\,{1-{\tx^{+m}_{\R1}} \tx^{+m}_{\L2}\ov  1-{ \tx^{-m}_{\R1}}\tx^{-m}_{\L2}}\,,
\eal
as required by supersymmetry.

Then for particles from different spheres we get ($m_1=m_2=m=1/2$)
\bal
\label{eq:SXXmmbhh}
S^{m_1m_2}_{\cX\cX} (u_1,u_2)&=H^{m_1m_2}_{\cX\cX} (u_1,u_2)
{\sqrt{\tx_{\L1}^{+m_1}}\ov \sqrt{\tx_{\L1}^{-m_2}}}\, {\sqrt{\tx_{\L2}^{-m_1}}\ov \sqrt{\tx_{\L2}^{+m_2}}}\, { \tx^{-m_1}_{\L1}- \tx^{+m_2}_{\L2}\ov \tx^{+m_1}_{\L1}- \tx^{-m_2}_{\L2}}
    \frac{u_{12} + {i\ov h}}{u_{12} - {i\ov h}}\,
    \\
    &\quad\times
    \frac{R(\tg^{+m_1+m_2}_{\L\L}) R(\tg^{-m_1-m_2}_{\L\L})}{R(\tg^{+m_1-m_2}_{\L\L}) R(\tg^{-m_1+m_2}_{\L\L}) } \left(\frac{\Sigma^{\bes}_{\L\L}(\tx^{\pm m_1}_{\L1}, \tx^{\pm m_2}_{\L2})}{\Sigma^{\hl}_{\L\L}(\tx^{\pm m_1}_{\L1}, \tx^{\pm m_2}_{\L2})}\right)^{-2}\,,
\\
 S^{m_1m_2}_{\bar{\Upsilon}\mathcal{X}} (u_1,u_2)&=H^{m_1m_2}_{\bar{\Upsilon}\mathcal{X}}
(u_1,u_2)\, {\sqrt{\tilde\a_{\L2}^{+m_2}}\ov \sqrt{\tilde\a_{\L2}^{-m_2}}}
{\sqrt{\tx_{\R1}^{+m_1}}\ov \sqrt{\tx_{\R1}^{-m_1}}}\, {\sqrt{\tx_{\L2}^{+m_2}}\ov \sqrt{\tx_{\L2}^{-m_2}}}\, { {1\ov \tx^{+m_1}_{\R1}}- \tx^{-m_2}_{\L2}\ov {1\ov \tx^{-m_1}_{\R1}}- \tx^{+m_2}_{\L2}}
 \\
    &\quad\times
\,\frac{R(\tg^{-m_1+m_2}_{ \R\L}+ i \pi) R(\tg^{+m_1-m_2}_{ \R\L}- i \pi)}{R(\tg^{-m_1-m_2}_{ \R\L}+ i \pi) R(\tg^{+m_1+m_2}_{\R\L}- i \pi)}  \left(\frac{\Sigma^{\bes}_{\R\L}(\tx^{\pm m_1}_{\R1}, \tx^{\pm m_2}_{\L2})}{\Sigma^{\hl}_{\R\L}(\tx^{\pm m_1}_{\R1}, \tx^{\pm m_2}_{\L2})}\right)^{-2}\,,
\eal
and the elements $S^{m_1m_2}_{{\Upsilon}{\Upsilon}} $ and $S^{m_1m_2}_{\bar\cX\Upsilon}(u_1,u_2)$ can be restored from the ones above 
\bal
\label{eq:SYYm1m2hh}
S^{m_1m_2}_{{\Upsilon}{\Upsilon}} (u_1,u_2)=&-H^{m_1m_2}_{\cX\cX} (u_1,u_2)\,  \frac{u_{12} + {i\ov h}}{u_{12} - {i\ov h}}\,
    \\
    &\quad\times
    \frac{R(\tg^{+m_1+m_2}_{\L\L}) R(\tg^{-m_1-m_2}_{\L\L})}{R(\tg^{+m_1-m_2}_{\L\L}) R(\tg^{-m_1+m_2}_{\L\L}) } \left(\frac{\Sigma^{\bes}_{\L\L}(\tx^{\pm m_1}_{\L1}, \tx^{\pm m_2}_{\L2})}{\Sigma^{\hl}_{\L\L}(\tx^{\pm m_1}_{\L1}, \tx^{\pm m_2}_{\L2})}\right)^{-2}
    \,,\\
S^{m_1m_2}_{\bar\cX\Upsilon}(u_1,u_2)=&H^{m_1m_2}_{\bar{\Upsilon}\mathcal{X}}
(u_1,u_2)\, {\sqrt{\tilde\a_{\L2}^{-m_2}}\ov \sqrt{\tilde\a_{\L2}^{+m_2}}}\,
\\
&\times\frac{R(\tg^{-m_1+m_2}_{ \R\L}- i \pi) R(\tg^{+m_1-m_2}_{ \R\L}+ i \pi)}{R(\tg^{-m_1-m_2}_{ \R\L}- i \pi) R(\tg^{+m_1+m_2}_{\R\L}+ i \pi)} \left(\frac{\Sigma^{\bes}_{\R\L}(\tx^{\pm m_1}_{\R1}, \tx^{\pm m_2}_{\L2})}{\Sigma^{\hl}_{\R\L}(\tx^{\pm m_1}_{\R1}, \tx^{\pm m_2}_{\L2})}\right)^{-2} \,.
\eal

\section{Comparison with the QSC proposal}
\label{app:scomparison}

In this appendix we consider the pure RR case of equal masses so that $m_1=m_2=m=1/2$, and $k=0$. We want to compare our proposal with the one in \cite{Cavaglia:2025icd,Chernikov:2025jko}. For definiteness we use the results in paper \cite{Cavaglia:2025icd} which according to its authors are equivalent to those in \cite{Chernikov:2025jko}. Since in the RR case there is no difference between left and right particles we drop the subscripts, and  use  the notation
\bal
\tx^{\pm m}_{\L} = \tx^{\pm m}_{\R} \equiv \tx^{\pm}\,,\quad x^{\pm m}_{\L} = x^{\pm m}_{\R} \equiv x^{\pm}\,,\quad m=1/2\,.
\eal
From the Bethe equations in \cite{Cavaglia:2025icd} we can extract the corresponding S-matrix elements up to factors of $x_1^+/x_1^-\, x_2^-/x_2^+$ which are difficult to fix because the level-matching condition is imposed in the QSC consideration. The S-matrix elements we are interested in are\footnote{We thank the authors of \cite{Cavaglia:2025icd} for providing us with those S-matrix elements.} 
\bal
S_{ss}(x_1^\pm,x_2^\pm)&= {u_1-u_2-{i\ov h}\ov u_1-u_2+{i\ov h}} \,{1\ov \sigma^{\bes}(x_1^\pm,x_2^\pm)\Sigma^{\rm extra}(x_1^\pm,x_2^\pm)\Sigma_{\rm new}(x_1^\pm,x_2^\pm)}\,,
\\
S_{s\check s}(x_1^\pm,x_2^\pm)&= {\Sigma_{\rm new}(x_1^\pm,x_2^\pm)\ov \sigma^{\bes}(x_1^\pm,x_2^\pm)\Sigma^{\rm extra}(x_1^\pm,x_2^\pm)}\,,
\eal
where $\sigma^{\bes}(x_1^\pm,x_2^\pm)$ is the usual BES phase which was denoted as $\Sigma_{\rm BES}(x_1^\pm,x_2^\pm)$ in \cite{Cavaglia:2025icd}. They correspond to our $S^{m m}_{\cX\cX}$ and $S^{m_1m_2}_{\cX\cX}$ elements, and in the symmetric sector we need to consider their product. We get
\bal
S_{ss}(x_1^\pm,x_2^\pm)S_{s\check s}(x_1^\pm,x_2^\pm)&= {u_1-u_2-{i\ov h}\ov u_1-u_2+{i\ov h}} \,{1\ov \sigma^{\bes}(x_1^\pm,x_2^\pm)^2\,\Sigma^{\rm extra}(x_1^\pm,x_2^\pm)^2}\,,
\eal
and by using \eqref{eq:SXXmm} and \eqref{eq:SXXm1m2}
\bal
   S^{m m}_{\cX\cX}(u_1,u_2)S^{m_1m_2}_{\cX\cX} (u_1,u_2)=&\left(H^{mm}_{\cX\cX}  (u_1,u_2)\right)^2\, {{\tx_{1}^{+}}\ov {\tx_{1}^{-}}}\, {{\tx_{2}^{-}}\ov {\tx_{2}^{+}}}\,\left({ \tx^{-}_{1}- \tx^{+}_{2}\ov \tx^{+}_{1}- \tx^{-}_{\L2}}\right)^2
    \frac{u_1-u_2 + {i\ov h}}{u_1-u_2 - {i\ov h}}\,
    \\
    &\quad\times
    \frac{R^2(\tg_{12}^{++})R^2(\tg_{12}^{--})} {R^2(\tg_{12}^{+-}) R^2(\tg_{12}^{-+})}
    \left(\frac{\Sigma^{\bes}(\tx^{\pm }_{1}, \tx^{\pm }_{2})}{\Sigma^{\hl}(\tx^{\pm }_{1}, \tx^{\pm }_{2})}\right)^{-2}
\,,
\eal
In the RR case the improved HL phase is expressed in terms of the $R$ functions  through the following relation \cite{Frolov:2025tda}
\bal
 \Sigma^\hl(\tx^{\pm}_1, \tx^{\pm }_2)^2=\frac{R(\tg_{12}^{++}+ i\pi) R(\tg_{12}^{++}-i \pi) R(\tg_{12}^{--}+ i\pi) R(\tg_{12}^{--}-i \pi)}{R(\tg_{12}^{+-}+ i\pi) R(\tg_{12}^{+-}-i \pi) R(\tg_{12}^{-+}+ i\pi) R(\tg_{12}^{-+}-i \pi)} \frac{R^2(\tg_{12}^{+-}) R^2(\tg_{12}^{-+})}{R^2(\tg_{12}^{++})R^2(\tg_{12}^{--})} \,.
\eal
Thus, in the RR case we get 
\bal
   S^{m m}_{\cX\cX}(u_1,u_2)S^{m_1m_2}_{\cX\cX} (u_1,u_2)&=\left(H^{mm}_{\cX\cX}  (u_1,u_2)\right)^2\, {{\tx_{1}^{+}}\ov {\tx_{1}^{-}}}\, {{\tx_{2}^{-}}\ov {\tx_{2}^{+}}}\,
    \frac{u_1-u_2 - {i\ov h}}{u_1-u_2 + {i\ov h}}\,
    \\
    &\times
  \frac{R(\tg_{12}^{++}+ i\pi) R(\tg_{12}^{++}-i \pi) R(\tg_{12}^{--}+ i\pi) R(\tg_{12}^{--}-i \pi)}{R(\tg_{12}^{+-}+ i\pi) R(\tg_{12}^{+-}-i \pi) R(\tg_{12}^{-+}+ i\pi) R(\tg_{12}^{-+}-i \pi)} 
      \\
    &\times
    \left({ 1-{1\ov \tx^{-}_{1} \tx^{+}_{2}}\ov 1-{1\ov \tx^{+}_{1} \tx^{-}_{2}}}{\Sigma^{\bes}(\tx^{\pm }_{1}, \tx^{\pm }_{2})}\right)^{-2}
\,,
\eal
where we use the factorisation $u_1-u_2 \pm {i\ov h} = (\tx_1^\pm-\tx_2^\mp) (1-{1\ov \tx_1^\pm\tx_2^\mp})$.
One can check numerically that
\bal
 \frac{R(\tg_{12}^{++}+ i\pi) R(\tg_{12}^{++}-i \pi) R(\tg_{12}^{--}+ i\pi) R(\tg_{12}^{--}-i \pi)}{R(\tg_{12}^{+-}+ i\pi) R(\tg_{12}^{+-}-i \pi) R(\tg_{12}^{-+}+ i\pi) R(\tg_{12}^{-+}-i \pi)}  = {1\ov \Sigma^{\rm extra}(\tx_1^\pm,\tx_2^\pm)^2}\,,
\eal
\bal
 \frac{R(\g_{12}^{++}+ i\pi) R(\g_{12}^{++}-i \pi) R(\g_{12}^{--}+ i\pi) R(\g_{12}^{--}-i \pi)}{R(\g_{12}^{+-}+ i\pi) R(\g_{12}^{+-}-i \pi) R(\g_{12}^{-+}+ i\pi) R(\g_{12}^{-+}-i \pi)}  = {1\ov \Sigma^{\rm extra}(x_1^\pm,x_2^\pm)^2}\,,
\eal
and, taking into account that  \cite{Frolov:2025uwz} 
\bal
{ 1-{1\ov \tx^{-}_{1} \tx^{+}_{2}}\ov 1-{1\ov \tx^{+}_{1} \tx^{-}_{2}}}{\Sigma^{\bes}(\tx^{\pm }_{1}, \tx^{\pm }_{2})} = \sigma^{\bes}(\tx^{\pm }_{1}, \tx^{\pm }_{2})\,,
\eal
 we get that up to the factor 
\bal
\left(H^{mm}_{\cX\cX}  (u_1,u_2)\right)^2\, {{\tx_{1}^{+}}\ov {\tx_{1}^{-}}}\, {{\tx_{2}^{-}}\ov {\tx_{2}^{+}}}\,,
\eal
the product of the S-matrix elements agrees with the one found in \cite{Cavaglia:2025icd,Chernikov:2025jko}. 

We also find the relation
\bal
 \frac{R(\tg_{12}^{++}) R(\tg_{12}^{--})}{R(\tg_{12}^{+-}) R(\tg_{12}^{-+})}  = { \widetilde{\Sigma}^{\rm extra}(\tx_1^\pm,\tx_2^\pm)^2}\,,\qquad  \frac{R(\g_{12}^{++}) R(\g_{12}^{--})}{R(\g_{12}^{+-}) R(\g_{12}^{-+})}  = { \widetilde{\Sigma}^{\rm extra}(x_1^\pm,x_2^\pm)^2}\,,
\eal
that can be used to check an agreement for the product of left-right S-matrix elements.

We get the same result for $S^{m m}_{\cX\cX}(u_1,u_2)S^{m_1m_2}_{\cX\cX} (u_1,u_2)$ if we use the proposal for the S-matrix elements in appendix \ref{app:solutions:2}. In fact, if we set the function e$(v)$ in \cite{Cavaglia:2025icd} to 0 then we can  easily solve the crossing equation (5.21) in \cite{Cavaglia:2025icd} for $\Sigma^{\rm new}$ by setting
\bal
\label{eq:oursigmanew}
\Sigma^{\rm new}(x_1^\pm,x_2^\pm)=\frac{\Sigma^{\hl}(x^{\pm }_{1}, x^{\pm }_{2})}{\Sigma^{\bes}(x^{\pm }_{1}, x^{\pm }_{2})}\,.
\eal
This solution, however, breaks some of the constraints which follow from the QSC construction (which are more stringent that crossing alone),
as those are incompatible with either the braiding unitarity or the
usual crossing equations: preserving the braiding unitarity requires  introducing a non-trivial function $e(v)$ in the crossing equations.
This notwithstanding, we see that using~\eqref{eq:oursigmanew}, the S-matrix element $S_{ss}$ of~\cite{Cavaglia:2025icd} would  become
\bal
S_{ss}(x^{\pm }_{1}, x^{\pm }_{2}) &= { x_1^- - x_2^+\ov x_1^+-x_2^-} {1\ov \Sigma^{\rm extra}(x_1^\pm,x_2^\pm )\Sigma^{\hl}(x_1^\pm,x_2^\pm)}
\\
&={x_1^+-x_2^-\ov x_1^- - x_2^+} \frac{R(\tg_{12}^{+-}) R(\tg_{12}^{-+})}{R(\tg_{12}^{++}) R(\tg_{12}^{--})} \,,
\eal
and up to the factor\footnote{Note that their S-matrix elements are inverted with respect to ours.}
\bal
H^{mm}_{\cX\cX}  (u_1,u_2)\,  {\sqrt{\tx_{\L1}^{+}}\ov \sqrt{\tx_{\L1}^{-}}}\, {\sqrt{\tx_{\L2}^{-}}\ov \sqrt{\tx_{\L2}^{+}}}\,,
  \eal
it is equal to $S^{m m}_{\cX\cX}(u_1,u_2)$ from appendix \ref{app:solutions:2}.  In this sense, the dressing factors proposed in \cite{Cavaglia:2025icd,Chernikov:2025jko} appear closer (but still not the same, especially since because their e$(v)$ is nontrivial) to ours from appendix \ref{app:solutions:2}.

\bibliographystyle{JHEP}
\bibliography{refs}

@article{Arutyunov:2009ax,
    author = "Arutyunov, Gleb and Frolov, Sergey and Suzuki, Ryo",
    title = "{Exploring the mirror TBA}",
    eprint = "0911.2224",
    archivePrefix = "arXiv",
    primaryClass = "hep-th",
    reportNumber = "ITP-UU-09-54, SPIN-09-44, TCDMATH-09-24, HMI-09-10",
    doi = "10.1007/JHEP05(2010)031",
    journal = "JHEP",
    volume = "05",
    pages = "031",
    year = "2010"
}

@article{Hofman:2006xt,
    author = "Hofman, Diego M. and Maldacena, Juan Martin",
    title = "{Giant Magnons}",
    eprint = "hep-th/0604135",
    archivePrefix = "arXiv",
    doi = "10.1088/0305-4470/39/41/S17",
    journal = "J. Phys. A",
    volume = "39",
    pages = "13095--13118",
    year = "2006"
}

@article{Witten:2024yod,
    author = "Witten, Edward",
    title = "{Instantons and the large ${\mathcal{N}} = 4$ algebra}",
    eprint = "2407.20964",
    archivePrefix = "arXiv",
    primaryClass = "hep-th",
    doi = "10.1088/1751-8121/ada64d",
    journal = "J. Phys. A",
    volume = "58",
    number = "3",
    pages = "035403",
    year = "2025"
}

@article{Eberhardt:2017pty,
    author = "Eberhardt, Lorenz and Gaberdiel, Matthias R. and Li, Wei",
    title = "{A holographic dual for string theory on AdS$_{3}${\texttimes}S$^{3}${\texttimes}S$^{3}${\texttimes}S$^{1}$}",
    eprint = "1707.02705",
    archivePrefix = "arXiv",
    primaryClass = "hep-th",
    doi = "10.1007/JHEP08(2017)111",
    journal = "JHEP",
    volume = "08",
    pages = "111",
    year = "2017"
}

@article{Borsato:2015mma,
    author = "Borsato, Riccardo and Ohlsson Sax, Olof and Sfondrini, Alessandro and Stefa{\'n}ski, Bogdan",
    title = "{The $\mathrm{AdS}_3\times \mathrm{S}^3\times \mathrm{S}^3\times\mathrm{S}^1$ worldsheet S matrix}",
    eprint = "1506.00218",
    archivePrefix = "arXiv",
    primaryClass = "hep-th",
    reportNumber = "ITP-UU-15-08, HU-EP-15-26, HU-MATHEMATIK-P-2015-06, IMPERIAL-TP-OOS-2015-01",
    doi = "10.1088/1751-8113/48/41/415401",
    journal = "J. Phys. A",
    volume = "48",
    number = "41",
    pages = "415401",
    year = "2015"
}

@article{Tong:2014yna,
    author = "Tong, David",
    title = "{The holographic dual of $AdS_{3} \times  S^{3} \times S^{3} \times S^{1}$}",
    eprint = "1402.5135",
    archivePrefix = "arXiv",
    primaryClass = "hep-th",
    doi = "10.1007/JHEP04(2014)193",
    journal = "JHEP",
    volume = "04",
    pages = "193",
    year = "2014"
}

@article{Cagnazzo:2012se,
    author = "Cagnazzo, A. and Zarembo, K.",
    title = "{B-field in AdS(3)/CFT(2) Correspondence and Integrability}",
    eprint = "1209.4049",
    archivePrefix = "arXiv",
    primaryClass = "hep-th",
    reportNumber = "NORDITA-2012-67, UUITP-24-12",
    doi = "10.1007/JHEP11(2012)133",
    journal = "JHEP",
    volume = "11",
    pages = "133",
    year = "2012",
    note = "[Erratum: JHEP 04, 003 (2013)]"
}

@article{Babichenko:2009dk,
    author = "Babichenko, A. and Stefanski, Jr., B. and Zarembo, K.",
    title = "{Integrability and the AdS(3)/CFT(2) correspondence}",
    eprint = "0912.1723",
    archivePrefix = "arXiv",
    primaryClass = "hep-th",
    reportNumber = "ITEP-TH-59-09, LPTENS-09-36, UUITP-25-09",
    doi = "10.1007/JHEP03(2010)058",
    journal = "JHEP",
    volume = "03",
    pages = "058",
    year = "2010"
}

@article{Elitzur:1998mm,
    author = "Elitzur, Shmuel and Feinerman, Ofer and Giveon, Amit and Tsabar, David",
    title = "{String theory on AdS(3) x S**3 x S**3 x S**1}",
    eprint = "hep-th/9811245",
    archivePrefix = "arXiv",
    reportNumber = "RI-11-98",
    doi = "10.1016/S0370-2693(99)00101-X",
    journal = "Phys. Lett. B",
    volume = "449",
    pages = "180--186",
    year = "1999"
}

@article{Maldacena:1997re,
    author = "Maldacena, Juan Martin",
    title = "{The Large $N$ limit of superconformal field theories and supergravity}",
    eprint = "hep-th/9711200",
    archivePrefix = "arXiv",
    reportNumber = "HUTP-97-A097, HUTP-98-A097",
    doi = "10.4310/ATMP.1998.v2.n2.a1",
    journal = "Adv. Theor. Math. Phys.",
    volume = "2",
    pages = "231--252",
    year = "1998"
}

@article{Boonstra:1998yu,
    author = "Boonstra, Harm Jan and Peeters, Bas and Skenderis, Kostas",
    title = "{Brane intersections, anti-de Sitter space-times and dual superconformal theories}",
    eprint = "hep-th/9803231",
    archivePrefix = "arXiv",
    reportNumber = "KUL-TF-98-17",
    doi = "10.1016/S0550-3213(98)00512-4",
    journal = "Nucl. Phys. B",
    volume = "533",
    pages = "127--162",
    year = "1998"
}

@article{Frappat:1996pb,
    author = "Frappat, L. and Sorba, P. and Sciarrino, A.",
    title = "{Dictionary on Lie superalgebras}",
    eprint = "hep-th/9607161",
    archivePrefix = "arXiv",
    reportNumber = "ENSLAPP-AL-600-96, DSF-T-30-96",
    month = "7",
    year = "1996"
}

@article{Maldacena:2000hw,
    author = "Maldacena, Juan Martin and Ooguri, Hirosi",
    title = "{Strings in AdS(3) and SL(2,R) WZW model 1.: The Spectrum}",
    eprint = "hep-th/0001053",
    archivePrefix = "arXiv",
    reportNumber = "CALT-68-2245, CITUSC-99-010, HUTP-99-A027, LBNL-44375, UCB-PTH-99-48, LBL-44375",
    doi = "10.1063/1.1377273",
    journal = "J. Math. Phys.",
    volume = "42",
    pages = "2929--2960",
    year = "2001"
}

@article{Sfondrini:2014via,
    author = "Sfondrini, Alessandro",
    title = "{Towards integrability for ${\rm Ad}{{{\rm S}}_{{\bf 3}}}/{\rm CF}{{{\rm T}}_{{\bf 2}}}$}",
    eprint = "1406.2971",
    archivePrefix = "arXiv",
    primaryClass = "hep-th",
    reportNumber = "HU-MATHEMATIK-2014-14, HU-EP-14-24",
    doi = "10.1088/1751-8113/48/2/023001",
    journal = "J. Phys. A",
    volume = "48",
    number = "2",
    pages = "023001",
    year = "2015"
}

@article{Demulder:2023bux,
    author = "Demulder, Saskia and Driezen, Sibylle and Knighton, Bob and Oling, Gerben and Retore, Ana L. and Seibold, Fiona K. and Sfondrini, Alessandro and Yan, Ziqi",
    title = "{Exact approaches on the string worldsheet}",
    eprint = "2312.12930",
    archivePrefix = "arXiv",
    primaryClass = "hep-th",
    reportNumber = "NORDITA 2023-083",
    doi = "10.1088/1751-8121/ad72be",
    journal = "J. Phys. A",
    volume = "57",
    number = "42",
    pages = "423001",
    year = "2024"
}

@article{Arutyunov:2009ga,
    author = "Arutyunov, Gleb and Frolov, Sergey",
    title = "{Foundations of the AdS$_{5} \times S^{5}$ Superstring. Part I}",
    eprint = "0901.4937",
    archivePrefix = "arXiv",
    primaryClass = "hep-th",
    reportNumber = "ITP-UU-09-05, SPIN-09-05, TCD-MATH-09-06, HMI-09-03",
    doi = "10.1088/1751-8113/42/25/254003",
    journal = "J. Phys. A",
    volume = "42",
    pages = "254003",
    year = "2009"
}

@article{Beisert:2010jr,
    author = "Beisert, Niklas and others",
    title = "{Review of AdS/CFT Integrability: An Overview}",
    eprint = "1012.3982",
    archivePrefix = "arXiv",
    primaryClass = "hep-th",
    reportNumber = "AEI-2010-175, CERN-PH-TH-2010-306, HU-EP-10-87, HU-MATH-2010-22, KCL-MTH-10-10, UMTG-270, UUITP-41-10",
    doi = "10.1007/s11005-011-0529-2",
    journal = "Lett. Math. Phys.",
    volume = "99",
    pages = "3--32",
    year = "2012"
}

@article{Arutyunov:2005hd,
    author = "Arutyunov, Gleb and Frolov, Sergey",
    title = "{Uniform light-cone gauge for strings in AdS(5) x s**5: Solving SU(1|1) sector}",
    eprint = "hep-th/0510208",
    archivePrefix = "arXiv",
    reportNumber = "ITP-UU-05-47, SPIN-05-32, AEI-2005-160",
    doi = "10.1088/1126-6708/2006/01/055",
    journal = "JHEP",
    volume = "01",
    pages = "055",
    year = "2006"
}

@article{Rughoonauth:2012qd,
    author = "Rughoonauth, Nitin and Sundin, Per and Wulff, Linus",
    title = "{Near BMN dynamics of the AdS(3) x S(3) x S(3) x S(1) superstring}",
    eprint = "1204.4742",
    archivePrefix = "arXiv",
    primaryClass = "hep-th",
    reportNumber = "MIFPA-12-17",
    doi = "10.1007/JHEP07(2012)159",
    journal = "JHEP",
    volume = "07",
    pages = "159",
    year = "2012"
}

@article{Sundin:2013ypa,
    author = "Sundin, Per and Wulff, Linus",
    title = "{Worldsheet scattering in AdS(3)/CFT(2)}",
    eprint = "1302.5349",
    archivePrefix = "arXiv",
    primaryClass = "hep-th",
    reportNumber = "MIFPA-13-08",
    doi = "10.1007/JHEP07(2013)007",
    journal = "JHEP",
    volume = "07",
    pages = "007",
    year = "2013"
}

@article{Dei:2018yth,
    author = "Dei, Andrea and Gaberdiel, Matthias R. and Sfondrini, Alessandro",
    title = "{The plane-wave limit of ${\rm AdS}_3 \times {\rm S}^3 \times {\rm S}^3 \times {\rm S}^1$}",
    eprint = "1805.09154",
    archivePrefix = "arXiv",
    primaryClass = "hep-th",
    doi = "10.1007/JHEP08(2018)097",
    journal = "JHEP",
    volume = "08",
    pages = "097",
    year = "2018"
}

@article{Arutyunov:2006ak,
    author = "Arutyunov, Gleb and Frolov, Sergey and Plefka, Jan and Zamaklar, Marija",
    title = "{The Off-shell Symmetry Algebra of the Light-cone AdS(5) x S**5 Superstring}",
    eprint = "hep-th/0609157",
    archivePrefix = "arXiv",
    reportNumber = "AEI-2006-071, HU-EP-06-31, ITP-UU-06-39, SPIN-06-33, TCDMATH-06-13",
    doi = "10.1088/1751-8113/40/13/018",
    journal = "J. Phys. A",
    volume = "40",
    pages = "3583--3606",
    year = "2007"
}

@article{Arutyunov:2006yd,
    author = "Arutyunov, Gleb and Frolov, Sergey and Zamaklar, Marija",
    title = "{The Zamolodchikov-Faddeev algebra for AdS(5) x S**5 superstring}",
    eprint = "hep-th/0612229",
    archivePrefix = "arXiv",
    reportNumber = "AEI-2006-099, ITP-UU-06-58, SPIN-06-48, RCDMATH-06-18",
    doi = "10.1088/1126-6708/2007/04/002",
    journal = "JHEP",
    volume = "04",
    pages = "002",
    year = "2007"
}

@article{Borsato:2012ud,
    author = "Borsato, Riccardo and Ohlsson Sax, Olof and Sfondrini, Alessandro",
    title = "{A dynamic $\mathfrak{su}$(1|1)$^2$ S-matrix for AdS$_3$/CFT$_2$}",
    eprint = "1211.5119",
    archivePrefix = "arXiv",
    primaryClass = "hep-th",
    reportNumber = "ITP-UU-12-46, SPIN-12-43",
    doi = "10.1007/JHEP04(2013)113",
    journal = "JHEP",
    volume = "04",
    pages = "113",
    year = "2013"
}

@article{Hoare:2013lja,
    author = "Hoare, B. and Stepanchuk, A. and Tseytlin, A. A.",
    title = "{Giant magnon solution and dispersion relation in string theory in $AdS_3$x$S^3$x$T^4$ with mixed flux}",
    eprint = "1311.1794",
    archivePrefix = "arXiv",
    primaryClass = "hep-th",
    reportNumber = "IMPERIAL-TP-AS-2013-01, HU-EP-13-56",
    doi = "10.1016/j.nuclphysb.2013.12.011",
    journal = "Nucl. Phys. B",
    volume = "879",
    pages = "318--347",
    year = "2014"
}

@article{Dei:2018jyj,
    author = "Dei, Andrea and Sfondrini, Alessandro",
    title = "{Integrable S matrix, mirror TBA and spectrum for the stringy AdS$_{3}$ {\texttimes} S$^{3}$ {\texttimes} S$^{3}$ {\texttimes} S$^{1}$ WZW model}",
    eprint = "1812.08195",
    archivePrefix = "arXiv",
    primaryClass = "hep-th",
    doi = "10.1007/JHEP02(2019)072",
    journal = "JHEP",
    volume = "02",
    pages = "072",
    year = "2019"
}

@article{Zamolodchikov:1989cf,
    author = "Zamolodchikov, A. B.",
    title = "{Thermodynamic Bethe Ansatz in Relativistic Models. Scaling Three State Potts and Lee-yang Models}",
    reportNumber = "ITEP-89-144",
    doi = "10.1016/0550-3213(90)90333-9",
    journal = "Nucl. Phys. B",
    volume = "342",
    pages = "695--720",
    year = "1990"
}

@article{Arutyunov:2007tc,
    author = "Arutyunov, Gleb and Frolov, Sergey",
    title = "{On String S-matrix, Bound States and TBA}",
    eprint = "0710.1568",
    archivePrefix = "arXiv",
    primaryClass = "hep-th",
    reportNumber = "ITP-UU-07-50, SPIN-07-37, TCDMATH-07-15",
    doi = "10.1088/1126-6708/2007/12/024",
    journal = "JHEP",
    volume = "12",
    pages = "024",
    year = "2007"
}

@article{Dorey:1996re,
    author = "Dorey, Patrick and Tateo, Roberto",
    title = "{Excited states by analytic continuation of TBA equations}",
    eprint = "hep-th/9607167",
    archivePrefix = "arXiv",
    reportNumber = "DTP-96-29",
    doi = "10.1016/S0550-3213(96)00516-0",
    journal = "Nucl. Phys. B",
    volume = "482",
    pages = "639--659",
    year = "1996"
}

@article{Frolov:2025tda,
    author = "Frolov, Sergey and Polvara, Davide and Sfondrini, Alessandro",
    title = "{Dressing Factors and Mirror Thermodynamic Bethe Ansatz for mixed-flux AdS3/CFT2}",
    eprint = "2507.12191",
    archivePrefix = "arXiv",
    primaryClass = "hep-th",
    reportNumber = "ZMP-HH/25-12",
    month = "7",
    year = "2025"
}

@article{Frolov:2025uwz,
    author = "Frolov, Sergey and Polvara, Davide and Sfondrini, Alessandro",
    title = "{Massive dressing factors for mixed-flux AdS$_{3}$/CFT$_{2}$}",
    eprint = "2501.05995",
    archivePrefix = "arXiv",
    primaryClass = "hep-th",
    reportNumber = "ZMP-HH/25-1",
    doi = "10.1007/JHEP07(2025)171",
    journal = "JHEP",
    volume = "07",
    pages = "171",
    year = "2025"
}

@article{Cavaglia:2025icd,
    author = "Cavagli{\`a}, Andrea and Frassek, Rouven and Primi, Nicol{\`o} and Tateo, Roberto",
    title = "{On the Quantum Spectral Curve for $\text{AdS}_3\times \text{S}^3\times \text{S}^3\times \text{S}^1$ strings and the $\mathfrak{d}(2,1;\alpha)$ Q-system}",
    eprint = "2511.09635",
    archivePrefix = "arXiv",
    primaryClass = "hep-th",
    month = "11",
    year = "2025"
}

@article{Chernikov:2025jko,
    author = "Chernikov, Filipp and Ekhammar, Simon and Gromov, Nikolay and Smith, Benjamin",
    title = "{Gluing Quantum Spectral Curves: A Two-Copy osp(4|2) Construction}",
    eprint = "2511.09654",
    archivePrefix = "arXiv",
    primaryClass = "hep-th",
    month = "11",
    year = "2025"
}

@article{Sundin:2012gc,
    author = "Sundin, Per and Wulff, Linus",
    title = "{Classical integrability and quantum aspects of the AdS(3) x S(3) x S(3) x S(1) superstring}",
    eprint = "1207.5531",
    archivePrefix = "arXiv",
    primaryClass = "hep-th",
    reportNumber = "MIFPA-12-26",
    doi = "10.1007/JHEP10(2012)109",
    journal = "JHEP",
    volume = "10",
    pages = "109",
    year = "2012"
}

@article{Zarembo:2009au,
    author = "Zarembo, K.",
    title = "{Worldsheet spectrum in AdS(4)/CFT(3) correspondence}",
    eprint = "0903.1747",
    archivePrefix = "arXiv",
    primaryClass = "hep-th",
    reportNumber = "ITEP-TH-11-09, LPTENS-09-05, UUITP-08-09",
    doi = "10.1088/1126-6708/2009/04/135",
    journal = "JHEP",
    volume = "04",
    pages = "135",
    year = "2009"
}

@article{Lloyd:2014bsa,
    author = "Lloyd, Thomas and Ohlsson Sax, Olof and Sfondrini, Alessandro and Stefa{\'n}ski, Jr., Bogdan",
    title = "{The complete worldsheet S matrix of superstrings on AdS$_3 \times$ S$^3 \times$ T$^4$ with mixed three-form flux}",
    eprint = "1410.0866",
    archivePrefix = "arXiv",
    primaryClass = "hep-th",
    reportNumber = "IMPERIAL-TP-OOS-2014-04, HU-MATHEMATIK-2014-21, HU-EP-14-34",
    doi = "10.1016/j.nuclphysb.2014.12.019",
    journal = "Nucl. Phys. B",
    volume = "891",
    pages = "570--612",
    year = "2015"
}

@article{Bombardelli:2009xz,
    author = "Bombardelli, Diego and Fioravanti, Davide and Tateo, Roberto",
    title = "{TBA and Y-system for planar AdS(4)/CFT(3)}",
    eprint = "0912.4715",
    archivePrefix = "arXiv",
    primaryClass = "hep-th",
    doi = "10.1016/j.nuclphysb.2010.04.005",
    journal = "Nucl. Phys. B",
    volume = "834",
    pages = "543--561",
    year = "2010"
}

@article{Beisert:2006ez,
    author = "Beisert, Niklas and Eden, Burkhard and Staudacher, Matthias",
    title = "{Transcendentality and Crossing}",
    eprint = "hep-th/0610251",
    archivePrefix = "arXiv",
    reportNumber = "AEI-2006-079, ITP-UU-06-44, SPIN-06-34",
    doi = "10.1088/1742-5468/2007/01/P01021",
    journal = "J. Stat. Mech.",
    volume = "0701",
    pages = "P01021",
    year = "2007"
}

@article{Dorey:2007xn,
    author = "Dorey, Nick and Hofman, Diego M. and Maldacena, Juan Martin",
    title = "{On the Singularities of the Magnon S-matrix}",
    eprint = "hep-th/0703104",
    archivePrefix = "arXiv",
    doi = "10.1103/PhysRevD.76.025011",
    journal = "Phys. Rev. D",
    volume = "76",
    pages = "025011",
    year = "2007"
}

@article{Hernandez:2006tk,
    author = "Hernandez, Rafael and Lopez, Esperanza",
    title = "{Quantum corrections to the string Bethe ansatz}",
    eprint = "hep-th/0603204",
    archivePrefix = "arXiv",
    reportNumber = "CERN-PH-TH-2006-048, IFT-UAM-CSIC-06-14",
    doi = "10.1088/1126-6708/2006/07/004",
    journal = "JHEP",
    volume = "07",
    pages = "004",
    year = "2006"
}

@article{Arutyunov:2004vx,
    author = "Arutyunov, Gleb and Frolov, Sergey and Staudacher, Matthias",
    title = "{Bethe ansatz for quantum strings}",
    eprint = "hep-th/0406256",
    archivePrefix = "arXiv",
    reportNumber = "AEI-2004-046",
    doi = "10.1088/1126-6708/2004/10/016",
    journal = "JHEP",
    volume = "10",
    pages = "016",
    year = "2004"
}

@article{Bianchi:2014rfa,
    author = "Bianchi, Lorenzo and Hoare, Ben",
    title = "{$AdS_3 \times S^3 \times M^4$ string S-matrices from unitarity cuts}",
    eprint = "1405.7947",
    archivePrefix = "arXiv",
    primaryClass = "hep-th",
    reportNumber = "HU-EP-14-18",
    doi = "10.1007/JHEP08(2014)097",
    journal = "JHEP",
    volume = "08",
    pages = "097",
    year = "2014"
}

@article{Frolov:2025ozz,
    author = "Frolov, Sergey and Polvara, Davide and Sfondrini, Alessandro",
    title = "{Exchange relations and crossing}",
    eprint = "2506.04096",
    archivePrefix = "arXiv",
    primaryClass = "hep-th",
    reportNumber = "ZMP-HH/25-9",
    doi = "10.1088/1751-8121/ae0edb",
    journal = "J. Phys. A",
    volume = "58",
    number = "41",
    pages = "415402",
    year = "2025"
}

@article{Colomo:1991gw,
    author = "Colomo, F. and Koubek, A. and Mussardo, G.",
    title = "{On the S matrix of the subleading magnetic deformation of the tricritical Ising model in two-dimensions}",
    eprint = "hep-th/9108024",
    archivePrefix = "arXiv",
    reportNumber = "NORDITA-91-47, SISSA-94-91-EP",
    doi = "10.1142/S0217751X92002416",
    journal = "Int. J. Mod. Phys. A",
    volume = "7",
    pages = "5281--5306",
    year = "1992"
}

@article{Frolov:2019nrr,
    author = "Frolov, Sergey A.",
    title = "{$T\overline T $ Deformation and the Light-Cone Gauge}",
    eprint = "1905.07946",
    archivePrefix = "arXiv",
    primaryClass = "hep-th",
    reportNumber = "TCD-MATH-19-06",
    doi = "10.1134/S0081543820030098",
    journal = "Proc. Steklov Inst. Math.",
    volume = "309",
    number = "1",
    pages = "107--126",
    year = "2020"
}

@article{Baggio:2018gct,
    author = "Baggio, Marco and Sfondrini, Alessandro",
    title = "{Strings on NS-NS Backgrounds as Integrable Deformations}",
    eprint = "1804.01998",
    archivePrefix = "arXiv",
    primaryClass = "hep-th",
    doi = "10.1103/PhysRevD.98.021902",
    journal = "Phys. Rev. D",
    volume = "98",
    number = "2",
    pages = "021902",
    year = "2018"
}

@article{Smirnov:2016lqw,
    author = "Smirnov, F. A. and Zamolodchikov, A. B.",
    title = "{On space of integrable quantum field theories}",
    eprint = "1608.05499",
    archivePrefix = "arXiv",
    primaryClass = "hep-th",
    doi = "10.1016/j.nuclphysb.2016.12.014",
    journal = "Nucl. Phys. B",
    volume = "915",
    pages = "363--383",
    year = "2017"
}

@article{Cavaglia:2016oda,
    author = "Cavagli{\`a}, Andrea and Negro, Stefano and Sz{\'e}cs{\'e}nyi, Istv{\'a}n M. and Tateo, Roberto",
    title = "{$T \bar{T}$-deformed 2D Quantum Field Theories}",
    eprint = "1608.05534",
    archivePrefix = "arXiv",
    primaryClass = "hep-th",
    doi = "10.1007/JHEP10(2016)112",
    journal = "JHEP",
    volume = "10",
    pages = "112",
    year = "2016"
}

@article{Arutyunov:2009kf,
    author = "Arutyunov, Gleb and Frolov, Sergey",
    title = "{The Dressing Factor and Crossing Equations}",
    eprint = "0904.4575",
    archivePrefix = "arXiv",
    primaryClass = "hep-th",
    reportNumber = "ITP-UU-09-17, SPIN-09-17, TCDMATH-09-12, HMI-09-06",
    doi = "10.1088/1751-8113/42/42/425401",
    journal = "J. Phys. A",
    volume = "42",
    pages = "425401",
    year = "2009"
}

@article{Seibold:2024qkh,
    author = "Seibold, Fiona K. and Sfondrini, Alessandro",
    title = "{AdS3 Integrability, Tensionless Limits, and Deformations: A Review}",
    eprint = "2408.08414",
    archivePrefix = "arXiv",
    primaryClass = "hep-th",
    month = "8",
    year = "2024"
}

@article{Stepanchuk:2014kza,
    author = "Stepanchuk, A.",
    title = "{String theory in $Ad{{S}_{3}}\times {{S}^{3}}\times {{T}^{4}}$ with mixed flux: semiclassical and 1-loop phase in the S-matrix}",
    eprint = "1412.4764",
    archivePrefix = "arXiv",
    primaryClass = "hep-th",
    reportNumber = "IMPERIAL-TP-AS-2014-01",
    doi = "10.1088/1751-8113/48/19/195401",
    journal = "J. Phys. A",
    volume = "48",
    number = "19",
    pages = "195401",
    year = "2015"
}

@article{Frolov:2023lwd,
    author = "Frolov, Sergey and Polvara, Davide and Sfondrini, Alessandro",
    title = "{On mixed-flux worldsheet scattering in AdS$_{3}$/CFT$_{2}$}",
    eprint = "2306.17553",
    archivePrefix = "arXiv",
    primaryClass = "hep-th",
    doi = "10.1007/JHEP11(2023)055",
    journal = "JHEP",
    volume = "11",
    pages = "055",
    year = "2023"
}

@article{Borsato:2012ss,
    author = "Borsato, Riccardo and Ohlsson Sax, Olof and Sfondrini, Alessandro",
    title = "{All-loop Bethe ansatz equations for AdS3/CFT2}",
    eprint = "1212.0505",
    archivePrefix = "arXiv",
    primaryClass = "hep-th",
    reportNumber = "ITP-UU-12-48, SPIN-12-45",
    doi = "10.1007/JHEP04(2013)116",
    journal = "JHEP",
    volume = "04",
    pages = "116",
    year = "2013"
}

@article{OhlssonSax:2011ms,
    author = "Ohlsson Sax, Olof and Stefanski, Jr., B.",
    title = "{Integrability, spin-chains and the AdS3/CFT2 correspondence}",
    eprint = "1106.2558",
    archivePrefix = "arXiv",
    primaryClass = "hep-th",
    reportNumber = "UUITP-17-11",
    doi = "10.1007/JHEP08(2011)029",
    journal = "JHEP",
    volume = "08",
    pages = "029",
    year = "2011"
}

@article{Dubovsky:2012wk,
    author = "Dubovsky, Sergei and Flauger, Raphael and Gorbenko, Victor",
    title = "{Solving the Simplest Theory of Quantum Gravity}",
    eprint = "1205.6805",
    archivePrefix = "arXiv",
    primaryClass = "hep-th",
    doi = "10.1007/JHEP09(2012)133",
    journal = "JHEP",
    volume = "09",
    pages = "133",
    year = "2012"
}

@article{Arutyunov:2011uz,
    author = "Arutyunov, Gleb and Frolov, Sergey",
    title = "{Comments on the Mirror TBA}",
    eprint = "1103.2708",
    archivePrefix = "arXiv",
    primaryClass = "hep-th",
    reportNumber = "ITP-UU-11-08, TCDMATH-11-04, HMI-11-03",
    doi = "10.1007/JHEP05(2011)082",
    journal = "JHEP",
    volume = "05",
    pages = "082",
    year = "2011"
}

@article{Beisert:2006ib,
    author = "Beisert, Niklas and Hernandez, Rafael and Lopez, Esperanza",
    title = "{A Crossing-symmetric phase for AdS(5)xS**5 strings}",
    eprint = "hep-th/0609044",
    archivePrefix = "arXiv",
    reportNumber = "AEI-2006-068, CERN-PH-TH-2006-176, IFT-UAM-CSIC-06-44, PUTP-2208",
    doi = "10.1088/1126-6708/2006/11/070",
    journal = "JHEP",
    volume = "11",
    pages = "070",
    year = "2006"
}

@article{Beisert:2005fw,
    author = "Beisert, Niklas and Staudacher, Matthias",
    title = "{Long-range $psu(2,2|4)$ Bethe Ansatze for gauge theory and strings}",
    eprint = "hep-th/0504190",
    archivePrefix = "arXiv",
    reportNumber = "AEI-2005-092, PUTP-2159",
    doi = "10.1016/j.nuclphysb.2005.06.038",
    journal = "Nucl. Phys. B",
    volume = "727",
    pages = "1--62",
    year = "2005"
}

\end{document}